\documentclass[12pt,letter]{article}
\usepackage{amsmath,amssymb,amsfonts, amsbsy, epsfig, subfig}
\usepackage[usenames]{color}
\usepackage{rotating}
\usepackage{setspace}

\newtheorem{theorem}{Theorem}
\newtheorem{corollary}{Corollary}
\newtheorem{proposition}{Proposition}
\newtheorem{lemma}{Lemma}
\newtheorem{example}{Example}
\newtheorem{definition}{Definition}
\newcommand{\beq}{\begin{equation}}
\newcommand{\eeq}{\end{equation}}
\newcommand{\beas}{\begin{eqnarray*}}
\newcommand{\eeas}{\end{eqnarray*}}
\newcommand{\bea}{\begin{eqnarray}}
\newcommand{\eea}{\end{eqnarray}}
\newcommand{\bei}{\begin{itemize}}
\newcommand{\eei}{\end{itemize}}
\newcommand{\ben}{\begin{enumerate}}
\newcommand{\een}{\end{enumerate}}
\newcommand{\bet}{\begin{theorem}}
\newcommand{\eet}{\end{theorem}}
\newcommand{\bel}{\begin{lemma}}
\newcommand{\eel}{\end{lemma}}
\newcommand{\bep}{\begin{proposition}}
\newcommand{\eep}{\end{proposition}}
\newcommand{\bed}{\begin{definition}}
\newcommand{\eed}{\end{definition}}
\newcommand{\bec}{\begin{corollary}}
\newcommand{\eec}{\end{corollary}}
\newcommand{\bex}{\begin{example}}
\newcommand{\eex}{\end{example}}
\newcommand{\qed}{\quad\hbox{\vrule width 4pt height 6pt depth 1.5pt}}
\newcommand{\RR}{I\!\! R}

\newcommand{\ep}{\epsilon}

\newcommand{\argmin}{\mathop{\rm arg\min}}
\newcommand{\argmax}{\mathop{\rm arg\max}}
\newcommand{\sgn}{\mathop{\rm sgn}\nolimits}

\def\pr{\textsf{P}} 
\def\ep{\textsf{E}} 

\begin{document}

\title{A Constrained $\ell_1$ Minimization Approach to Sparse Precision Matrix Estimation}
\author{Tony Cai$^{1}$, Weidong Liu$^{1, 2}$ and Xi Luo$^{1}$}
\date{}

\maketitle

\footnotetext[1]{ Department of Statistics, The Wharton School, University of
  Pennsylvania,  Philadelphia, PA  \newline  \indent \ \
19104, tcai@wharton.upenn.edu. The research of Tony Cai was supported in part
by NSF FRG   \newline  \indent \ \
Grant DMS-0854973.}
\footnotetext[2]{Shanghai Jiao Tong University, Shanghai, China.}

\begin{abstract}
A constrained $\ell_1$ minimization method is proposed for
estimating a sparse inverse covariance matrix
based on a sample of $n$ iid $p$-variate random variables.
The resulting estimator is shown to enjoy a
number of desirable properties. In particular, it is shown that the
rate of convergence between the estimator and the true $s$-sparse
precision matrix under the spectral norm is $s\sqrt{\log p/n}$ when
the population distribution has either exponential-type tails or
polynomial-type tails. Convergence rates under the elementwise
$\ell_{\infty}$ norm and Frobenius norm are also presented.
In addition, graphical model selection is considered.
The procedure is easily implementable by linear programming.
Numerical performance of the estimator
is investigated using both simulated and real data. In particular, the
procedure is applied to analyze a breast cancer dataset.
The procedure performs favorably in comparison to existing methods.
\end{abstract}

\noindent{\bf Keywords:\/} constrained $\ell_1$ minimization,
covariance matrix, Frobenius norm, Gaussian graphical model, rate of
convergence, precision matrix, spectral norm.


\newpage

\section{Introduction}
\setcounter{equation}{0}

Estimation of covariance matrix and its inverse is an important problem in
many areas of statistical analysis. Among many interesting examples
are principal component analysis, linear/quadratic discriminant
analysis, and graphical models.
Stable and accurate covariance estimation is becoming increasingly
more important in the high dimensional setting where the dimension $p$
can be much larger than the sample size $n$. In this setting classical
methods and results based on fixed $p$ and large $n$ are no longer applicable.
An additional challenge in the high dimensional setting is the
computational costs. It is important that estimation procedures are
computationally effective so that they can be used in high dimensional
applications.

Let $\textbf{X}=(X_{1},\dotsc, X_{p})$ be a $p$-variate random
vector with covariance matrix $\boldsymbol{\Sigma}_{0}$ and precision matrix
$\boldsymbol{\Omega}_{0}:=\boldsymbol{\Sigma}_{0}^{-1}$. Given an independent and identically
distributed random sample $\{\textbf{X}_{1},\dotsc,\textbf{X}_{n}\}$
from the distribution of $\textbf{X}$,  the most natural estimator
of $\boldsymbol{\Sigma}_{0}$ is perhaps
\begin{eqnarray*}
  \boldsymbol{\Sigma}_{n}=\frac{1}{n}\sum_{k=1}^{n}(\textbf{X}_{k}-\bar{\textbf{X}})(\textbf{X}_{k}-\bar{\textbf{X}})^{T},
\end{eqnarray*}
where $\bar{\textbf{X}}=n^{-1}\sum_{k=1}^{n}\textbf{X}_{k}$.  However, $\boldsymbol{\Sigma}_n$ is singular if
$p> n$, and thus is unstable for estimating $\boldsymbol{\Sigma}_0$, not to
mention that one cannot use its inverse to estimate the precision
matrix $\boldsymbol{\Omega}_0$.  In
order to estimate the covariance matrix $\boldsymbol{\Sigma}_{0}$ consistently, special
structures are usually imposed and various
estimators have been introduced under these assumptions.  When the
variables exhibit a certain ordering structure, which is often
the case for time series data, Bickel and Levina (2008a) proved that
banding the sample covariance matrix leads to a consistent
estimator. Cai, Zhang and Zhou (2010) established the minimax rate of
convergence and introduced a rate-optimal tapering estimator.
El Karoui (2008) and Bickel and Levina (2008b) proposed thresholding of the sample
covariance matrix for estimating a class of sparse covariance
matrices and obtained rates of convergence for the thresholding estimators.

Estimation of the precision matrix $\boldsymbol{\Omega}_{0}$ is more involved due
to the lack of a natural pivotal estimator like $\boldsymbol{\Sigma}_n$.
Assuming certain ordering structures,  methods based on banding the Cholesky
factor of the inverse have been proposed and studied. See, e.g.,  Wu and
Pourahmadi (2003), Huang et al.\ (2006), Bickel and Levina
(2008b).
Penalized likelihood methods have also been introduced for estimating sparse precision matrices.
In particular, the $\ell_1$ penalized normal likelihood estimator and its
variants, which shall be called $\ell_1$-MLE type estimators, were
considered in
several papers; see, for example, Yuan and Lin (2007), Friedman et al.\
(2008), d'Aspremont et al.\  (2008), and Rothman et al.\ (2008).
Convergence rate under the Frobenius norm loss was given in
Rothman et al.\ (2008).  Yuan (2009) derived the rates
of convergence for subgaussian distributions.
Under more restrictive conditions such as mutual
incoherence or irrepresentable conditions, Ravikumar et al.\ (2008)
obtained the rates of convergence in the elementwise $\ell_{\infty}$
norm and spectral norm.  Nonconvex penalties, usually computationally
more demanding, have also been considered under the same normal
likelihood model.  For example, Lam and Fan (2009) and Fan et al.\ (2009)
considered penalizing the normal likelihood with the nonconvex SCAD
penalty.  The main goal is to ameliorate
the bias problem due to $\ell_{1}$ penalization.

A closely related problem is the recovery of the support of the precision matrix,
 which is strongly connected to the selection of graphical models.  To be more specific, let $G=(V,E)$ be a
graph representing conditional independence relations between components of $\textbf{X}$. The vertex set $V$ has $p$ components $X_{1},\dotsc,
X_{p}$ and the edge set $E$ consists of ordered pairs $ (i,j)$, where
$(i,j)\in E$ if there is an edge between $X_{i}$ and $X_{j}$.  The
edge between $X_{i}$ and $X_{j}$ is excluded from $E$ if and only if $X_{i}$ and $X_{j}$ are independent given $(X_{k},k\neq i,j)$.  If $\textbf{X}\sim
N(\boldsymbol{\mu}_{0},\boldsymbol{\Sigma}_{0})$, then the conditional independence between $X_{i}$ and $X_{j}$ given other variables is equivalent to $\omega^{0}_{ij}=0$,
where we set $\boldsymbol{\Omega}_{0}=(\omega^{0}_{ij})$. Hence, for Gaussian
distributions, recovering the structure of the graph $G$ is equivalent to the estimation of the
support of the precision matrix (Lauritzen (1996)). A recent paper by Liu et al.\ (2009) showed that for a class of non-Gaussian distribution called
nonparanormal distribution, the problem of estimating the graph can also be reduced to the estimation of the precision matrix. In an important paper, Meinshausen and B\"{u}hlmann (2006) demonstrated convincingly a neighborhood selection approach to recover the support of
$\boldsymbol{\Omega}_0$ in a row by row fashion.
Yuan (2009) replaced the lasso selection  by a Dantzig type modification, where first the ratios between the off-diagonal elements $\omega_{ij}$ and
the corresponding diagonal element $\omega_{ii}$ were estimated for each row $i$ and then the diagonal entries $\omega_{ii}$ were obtained given the estimated ratios.  Convergence rates under the matrix
$\ell_1$ norm and spectral norm losses were established.

In the present paper, we study estimation of the precision matrix
$\boldsymbol{\Omega}_0$ for both sparse and non-sparse matrices, without restricting
to a specific sparsity pattern. In addition, graphical model selection
is also considered. A new method of constrained $\ell_1$-minimization
for inverse matrix estimation (CLIME) is introduced. Rates of
convergence in spectral norm as well as elementwise $\ell_{\infty}$
norm and Frobenius norm are established under weaker assumptions, and
are shown to be faster than those given for the $\ell_1$-MLE
estimators when the population distribution has polynomial-type
tails. A matrix is called $s$-sparse if there are at most $s$ non-zero
elements on each row. It is shown that when $\boldsymbol{\Omega}_{0}$ is $s$-sparse
and $\textbf{X}$ has either exponential-type or polynomial-type tails,
the error between our estimator $\hat{\boldsymbol{\Omega}}$ and $\boldsymbol{\Omega}_{0}$
satisfies $\|\hat{\boldsymbol{\Omega}}-\boldsymbol{\Omega}_{0}\|_{2}=O_{\pr}(s\sqrt{\log p/n})$
and $|\hat{\boldsymbol{\Omega}}-\boldsymbol{\Omega}_{0}|_{\infty}=O_{\pr}(\sqrt{\log p/n})$,
where $\|\cdot\|_{2}$ and $|\cdot|_{\infty}$ are the spectral norm and
elementwise $l_{\infty}$ norm respectively. Properties of the CLIME
estimator for estimating banded precision matrices are also discussed.
The CLIME method can also be adopted for the selection of graphical
models, with an additional thresholding step. The elementwise
$\ell_\infty$ norm result is instrumental for graphical model selection.

In addition to its desirable theoretical properties, the CLIME
estimator is computationally very attractive for high dimensional
data. It can be obtained one column at a time by solving a linear
program, and the resulting matrix
estimator is formed by combining the vector solutions (after a simple
symmetrization). No outer iterations are needed  and the algorithm is
easily scalable. An R package of our method has been developed and
is publicly available on the web.
Numerical performance of the estimator is investigated using both
simulated and real data. In particular, the procedure is applied to
analyze a breast cancer dataset. Results show that the
procedure performs favorably in comparison to existing methods.

The rest of the paper is organized as follows. In
Section \ref{sec:estimation}, after basic notations and definitions are
introduced, we present the CLIME estimator. Theoretical properties
including the rates of convergence are established in Section \ref{sec:rate}.
Graphical model selection is discussed in Section \ref{sec:consistency}.
Numerical performance of the CLIME estimator is considered in
Section \ref{sec:simu} through simulation studies and a real
data analysis. Further discussions on the connections and differences of
our results with other related work are given in Section
\ref{sec:conclusion}.  The proofs of the main results are given in
Section \ref{sec:proof}.

\section{Estimation via Constrained $\ell_1$ Minimization}
\label{sec:estimation}

In compressed sensing and high dimensional linear regression
literature, it is now well understood that constrained
$\ell_1$ minimization provides an effective way for reconstructing a
sparse signal. See, for example, Donoho et al.\ (2006)
and Cand\`es and Tao (2007). A particularly simple and elementary
analysis of constrained $\ell_1$ minimization methods is given in Cai, Wang and Xu (2010).

In this section, we introduce a method of constrained $\ell_1$ minimization
for inverse covariance matrix estimation.
We begin with basic notations and definitions.
Throughout, for a vector
$\textbf{a}=(a_{1},\dotsc,a_{p})^{T}\in \RR^{p}$, define
$|\textbf{a}|_{1}=\sum_{j=1}^{p}|a_{j}|$ and
$|\textbf{a}|_{2}=\sqrt{\sum_{j=1}^{p}a^{2}_{j}}$. For a matrix
$\boldsymbol{A}=(a_{ij})\in\RR^{p\times q}$, we define the elementwise $l_{\infty}$
norm $|\boldsymbol{A}|_{\infty}=\max_{1\leq i\leq p,1\leq j\leq q}|a_{ij}|$, the
spectral norm $\|\boldsymbol{A}\|_{2}=\sup_{|\textbf{x}|_{2}\leq
  1}|\boldsymbol{A}\textbf{x}|_{2}$, the matrix $\ell_1$ norm
$\|\boldsymbol{A} \|_{L_{1}}=\max_{1\leq
  j\leq q}\sum_{i=1}^{p}|a_{ij}|$, the Frobenius norm
$\|\boldsymbol{A}\|_{F}=\sqrt{\sum_{i,j}a^{2}_{ij}}$, and the elementwise $\ell_1$ norm
$\|\boldsymbol{A}\|_{1}=\sum_{i=1}^{p}\sum_{j=1}^{q}|a_{i,j}|$.  $\boldsymbol{I}$ denotes a
$p\times p$ identity matrix. For any two index sets $T$ and $T^{'}$
and matrix $\boldsymbol{A}$, we use $\boldsymbol{A}_{TT^{'}}$ to denote the $|T|\times|T^{'}|$
matrix with rows and columns of $\boldsymbol{A}$ indexed by $T$ and $T^{'}$
respectively. The notation $\boldsymbol{A} \succ 0$ means that $\boldsymbol{A}$ is positive
definite.

\medskip
We now define our CLIME estimator. Let
$\{\hat{\boldsymbol{\Omega}}_{1}\}$ be the solution set of the following
optimization problem:
\begin{eqnarray}\label{c1}
  \min\|\boldsymbol{\Omega}\|_{1} ~~\mbox{subject to:}~~ |\boldsymbol{\Sigma}_{n}\boldsymbol{\Omega}-\boldsymbol{I}|_{\infty}\leq
  \lambda_{n},~~\boldsymbol{\Omega}\in \RR^{p\times p},
\end{eqnarray}
where $\lambda_{n}$ is a tuning parameter. In (\ref{c1}), we do not
impose the symmetry condition on $\boldsymbol{\Omega}$ and as a result the solution
is not symmetric in general. The final  CLIME estimator of $\boldsymbol{\Omega}_{0}$ is
obtained by symmetrizing $\hat{\boldsymbol{\Omega}}_{1}$ as follows. Write
$\hat{\boldsymbol{\Omega}}_{1}=(\hat{\omega}^{1}_{ij})=(\hat{\boldsymbol{\omega}}^{1}_{1},\dotsc,
\hat{\boldsymbol{\omega}}^{1}_{p})$. The CLIME estimator $\hat{\boldsymbol{\Omega}}$ of $\boldsymbol{\Omega}_{0}$
is defined as
\begin{eqnarray}\label{me1}
  \hat{\boldsymbol{\Omega}}=(\hat{\omega}_{ij}),~~\mbox{where~~}
  \hat{\omega}_{ij}=\hat{\omega}_{ji}=
  \hat{\omega}^{1}_{ij}I\{|\hat{\omega}^{1}_{ij}|\leq|\hat{\omega}^{1}_{ji}|\}+
  \hat{\omega}^{1}_{ji}I\{|\hat{\omega}^{1}_{ij}|>|\hat{\omega}^{1}_{ji}|\}.
\end{eqnarray}
In other words, between $\hat{\omega}^{1}_{ij}$ and
$\hat{\omega}^{1}_{ji}$, we take the one with smaller magnitude. It is
clear that $\hat{\boldsymbol{\Omega}}$ is a symmetric matrix. Moreover, Theorem
\ref{thn-2} shows that it is positive definite with high probability.

The convex program (\ref{c1}) can be further decomposed into $p$
vector minimization problems. Let $\boldsymbol{e}_{i}$ be a standard unit vector in
$\RR^{p}$ with $1$ in the $i$-th coordinate and $0$ in all other
coordinates. For $1\leq i\leq p$, let $\hat{\boldsymbol{\bf \beta}}_{i}$ be the
solution of the following convex optimization problem
\begin{eqnarray}\label{o1}
  \min|\boldsymbol{\beta}|_{1} ~~\mbox{subject to}~~ |\boldsymbol{\Sigma}_{n}\boldsymbol{\beta} - \boldsymbol{e}_i|_{\infty}\leq
  \lambda_{n},
\end{eqnarray}
where $\boldsymbol{\beta}$ is a vector in $\RR^{p}$. The following lemma shows that
solving the optimization problem (\ref{c1}) is equivalent to solving
the $p$ optimization problems (\ref{o1}). That is,
$\{\hat{\boldsymbol{\Omega}}_{1}\}=\{\hat{\textbf{B}}\}:=\{(\hat{\boldsymbol{\beta}}_{1},\dotsc,
\hat{\boldsymbol{\beta}}_{p})\}$. This simple observation is useful both for
implementation and technical analysis.

\begin{lemma}
\label{le1}
Let $\{\hat{\boldsymbol{\Omega}}_{1}\}$ be the solution set of (\ref{c1}) and
let $\{\hat{\textbf{B}}\}:=\{(\hat{\boldsymbol{\beta}}_{1},\dotsc, \hat{\boldsymbol{\beta}}_{p})\}$ where
$\hat{\boldsymbol{\beta}}_i$ are solutions to (\ref{o1}) for $i=1, ..., p$. Then
$\{\hat{\boldsymbol{\Omega}}_{1}\}=\{\hat{\textbf{B}}\}$.
\end{lemma}

\begin{figure}[htb!]
  \centering
  \includegraphics[width=0.7\textwidth]{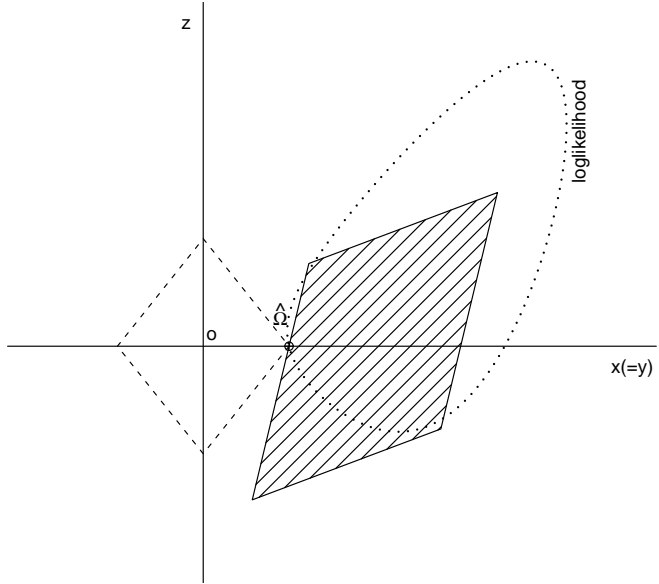}
  \caption{Plot of the elementwise $\ell_\infty$ constrained feasible set (
    shaded polygon) and the elementwise
    $\ell_1$ norm objective (dashed diamond near the origin) from
    CLIME.  The
    log-likelihood function as in Glasso is shown by the dotted line.}
  \label{fig:feasible}
\end{figure}

To illustrate the motivation of (\ref{c1}), let us recall the method
based on $\ell_{1}$ regularized log-determinant program (cf. d'Aspremont
et al.\ (2008), Friedman et al.\ (2008), Banerjee et al.\  (2008)) as
follows, which shall be called Glasso after  the algorithm that
efficiently computes the solution,
\begin{eqnarray}\label{glass}
  \hat{\boldsymbol{\Omega}}_{\rm Glasso}:=\argmin_{\Theta\succ
    0}\{\langle\boldsymbol{\Omega},\boldsymbol{\Sigma}_{n}\rangle-\log\det(\boldsymbol{\Omega})+\lambda_{n}\|\boldsymbol{\Omega}\|_{1}\}.
\end{eqnarray}
The solution $\hat{\boldsymbol{\Omega}}_{\rm Glasso}$ satisfies
\begin{eqnarray*}
  \hat{\boldsymbol{\Omega}}_{\rm Glasso}^{-1}-\boldsymbol{\Sigma}_{n}=\lambda_{n}\hat{\boldsymbol{Z}},
\end{eqnarray*}
where $\hat{\boldsymbol{Z}}$ is an element of the subdifferential
$\partial\|\hat{\boldsymbol{\Omega}}_{\rm Glasso}\|_{1}$. This leads us to consider
the optimization problem:
\begin{eqnarray}\label{c-c1}
  \min\|\boldsymbol{\Omega}\|_{1} ~~\mbox{subject to:}~~
  |\boldsymbol{\Omega}^{-1}-\boldsymbol{\Sigma}_{n}|_{\infty}\leq \lambda_{n},~~\boldsymbol{\Omega}\in
  \RR^{p\times p}.
\end{eqnarray}
However, the feasible set in (\ref{c-c1}) is very complicated.  By
multiplying the constraint with $\boldsymbol{\Omega}$,  such a  relaxation of
(\ref{c-c1}) leads to the convex optimization problem (\ref{c1}),
which can be easily solved.
Figure~\ref{fig:feasible} illustrates the
solution  for recovering a $2$ by $2$ precision  matrix $
[\begin{smallmatrix}
  x & z \\ z & y
\end{smallmatrix}]
$, and we only consider the plane  $x(= y)$ vs $z$ for simplicity.
The point where the feasible polygon  meets the dashed diamond is the
CLIME solution $\hat{\boldsymbol{\Omega}}$.  Note that the log-likelihood function as in Glasso is
a smooth curve as compared to the polygon constraint in  CLIME.

\section{Rates of Convergence}
\label{sec:rate}

In this section we investigate the theoretical properties of the CLIME
estimator and establish the rates of convergence
under different norms. Write
$\boldsymbol{\Sigma}_{n}=(\hat{\sigma}_{ij}) = (\hat{\sigma}_1, \dotsc, \hat{\sigma}_p)$, $\boldsymbol{\Sigma}_{0}=(\sigma^{0}_{ij})$ and $\ep
\textbf{X}=(\mu_{1},\dotsc,\mu_{p})$. It is conventional to divide the
technical analysis into two cases according to the moment conditions
on $\textbf{X}$.

{\bf (C1). (Exponential-type tails)} Suppose that there exists some
$0< \eta<1/4$ such that $\log p/n\leq \eta$ and
\begin{eqnarray*}
\ep e^{t(X_{i}-\mu_{i})^{2}}\leq K<\infty~~~\mbox{for all $|t|\leq
\eta$, for all $i$},
\end{eqnarray*}
where $K$ is a bounded constant.

{\bf (C2). (Polynomial-type tails)} Suppose that for some $\gamma,
c_{1}>0$,  $ p\leq c_{1}n^{\gamma}$, and for some $\delta>0$
\begin{eqnarray*}
\ep|X_{i}-\mu_{i}|^{4\gamma+4+\delta}\leq K~~~\mbox{for all $i$.}
\end{eqnarray*}

For $\ell_1$-MLE type estimators, it is typical that the convergence
rates in the case of polynomial-type tails are much slower than those
in the case of exponential-type tails. See, e.g., Ravikumar et al.\ (2008).
We shall show that our CLIME estimator attains the same rates of
convergence under either of the two moment conditions, and significantly
outperforms  $\ell_1$-MLE type estimators in the case of
polynomial-type tails.

\subsection{Rates of convergence under spectral norm}

We begin by considering the uniformity class of matrices:
\begin{eqnarray*}
  \mathcal{U}:=\mathcal{U}(q,s_{0}(p))=\Big{\{}\boldsymbol{\Omega}: \boldsymbol{\Omega}\succ
  0,~\|\boldsymbol{\Omega}\|_{L_{1}}\leq M,~\max_{1\leq i\leq
    p}\sum_{j=1}^{p}|\omega_{ij}|^{q}\leq s_{0}(p)\Big{\}}
\end{eqnarray*}
for $0\leq q<1$, where
$\boldsymbol{\Omega}=:(\omega_{ij})=(\boldsymbol{\omega}_{1},\dotsc,\boldsymbol{\omega}_{p})$. Similar parameter spaces have been used in Bickel and Levina (2008b) for estimating the covariance matrix $\boldsymbol{\Sigma}_{0}$ . Note that in the special case of $q=0$, $\mathcal{U}(0,s_{0}(p))$ is a class of $s_{0}(p)$-sparse matrices.
Let
\begin{eqnarray*}
  \theta=\max_{ij}\ep\Big{[}
  (X_{i}-\mu_{i})(X_{j}-\mu_{j})-\sigma^{0}_{ij}\Big{]}^{2}=:\max_{ij}\theta_{ij}.
\end{eqnarray*}
The quantity $\theta_{ij}$ is related to the variance of
$\hat{\sigma}_{ij}$, and the maximum value $\theta$ captures the
overall variability of $\boldsymbol{\Sigma}_n$.  It is easy to see that under either (C1) or (C2)
$\theta$ is a bounded constant depending only on $\gamma,\delta, K$.

The following theorem gives the rates of convergence for the CLIME estimator $\hat{\boldsymbol{\Omega}}$ under the spectral norm loss.
\begin{theorem}\label{thn-2}
  Suppose that $\boldsymbol{\Omega}_{0}\in\mathcal{U}(q,s_{0}(p))$.

  (i).  Assume (C1) holds. Let $\lambda_{n}=C_{0}M\sqrt{\log p/n}$,
  where $C_{0}=2\eta^{-2}(2+\tau+\eta^{-1}e^{2}K^{2})^{2}$ and
  $\tau>0$.  Then
  \begin{eqnarray}\label{d1}
    \|\hat{\boldsymbol{\Omega}}-\boldsymbol{\Omega}_{0}\|_{2}\leq C_{1}
    M^{2-2q}s_{0}(p)\Big{(}\frac{\log p}{n}\Big{)}^{(1-q)/2},
  \end{eqnarray}
  with probability greater than $1-4p^{-\tau}$, where $C_{1}\leq
  2(1+2^{1-q}+3^{1-q})4^{1-q}C_{0}^{1-q}$.

  (ii). Assume (C2) holds. Let $\lambda_{n}=C_{2}M\sqrt{\log p/n}$,
  where $C_{2}= \sqrt{(5+\tau)(\theta+1)}$. Then
  \begin{eqnarray}\label{d2}
   \|\hat{\boldsymbol{\Omega}}-\boldsymbol{\Omega}_{0}\|_{2}\leq C_{3}M^{2-2q}
    s_{0}(p)\Big{(}\frac{\log p}{n}\Big{)}^{(1-q)/2},
  \end{eqnarray}
  with probability greater than
  $1-O\Big{(}n^{-\delta/8}+p^{-\tau/2}\Big{)}$, where $C_{3}\leq
  2(1+2^{1-q}+3^{1-q})4^{1-q}C^{1-q}_{2}$.
\end{theorem}

When $M$ does not depend on $n,p$, the rates in Theorem \ref{thn-2}
are the same as those for estimating $\boldsymbol{\Sigma}_{0}$ in Bickel and Levina
(2008b). In the polynomial-type tails case and when $q=0$, the rate in
(\ref{d2}) is significantly better than the rate
$O\Big{(}s_{0}(p)\sqrt{\frac{p^{1/(\gamma+1+\delta/4)}}{n}}\Big{)}$
for the $\ell_1$-MLE estimator obtained in Ravikumar et al.\ (2008).

It would be of great interest to get the convergence rates for
$\sup_{\boldsymbol{\Omega}_{0}\in\mathcal{U}}\ep\|\hat{\boldsymbol{\Omega}}-\boldsymbol{\Omega}_{0}\|^{2}_{2}$.
However, it is even difficult to prove the existence of the
expectation of $\|\hat{\boldsymbol{\Omega}}-\boldsymbol{\Omega}_{0}\|^{2}_{2}$ as we are dealing
with the inverse matrix. We modify the estimator $\hat{\boldsymbol{\Omega}}$ to
ensure the existence of such expectation and the same rates are
established. Let $\{\hat{\boldsymbol{\Omega}}_{1\rho}\}$ be the solution set of the
following optimization problem:
\begin{eqnarray}\label{re-c1}
  \min\|\boldsymbol{\Omega}\|_{1} ~~\mbox{subj}~~
  |\boldsymbol{\Sigma}_{n,\rho}\boldsymbol{\Omega}-\boldsymbol{I}|_{\infty}\leq \lambda_{n},~~\boldsymbol{\Omega}\in
  \RR^{p\times p},
\end{eqnarray}
where $\boldsymbol{\Sigma}_{n,\rho}=\boldsymbol{\Sigma}_{n}+\rho \boldsymbol{I}$ with $\rho>0$.  Write
$\hat{\boldsymbol{\Omega}}_{1\rho}=(\hat{\omega}^{1}_{ij\rho})$. Define the
symmetrized estimator $\hat{\boldsymbol{\Omega}}_{\rho}$ as in (\ref{me1}) by
\begin{eqnarray}\label{me1-re}
  \hat{\boldsymbol{\Omega}}_{\rho}=(\hat{\omega}_{ij\rho}),~~\mbox{where~~}
  \hat{\omega}_{ij\rho}=\hat{\omega}_{ji\rho}=
  \hat{\omega}^{1}_{ij\rho}I\{|\hat{\omega}^{1}_{ij\rho}|\leq|\hat{\omega}^{1}_{ji\rho}|\}+
  \hat{\omega}^{1}_{ji\rho}I\{|\hat{\omega}^{1}_{ij\rho}|>|\hat{\omega}^{1}_{ji\rho}|\}.
\end{eqnarray}
Clearly $\boldsymbol{\Sigma}_{n,\rho}^{-1}$ is a feasible point, and thus we have
$\|\hat{\boldsymbol{\Omega}}_{1\rho}\|_{L_{1}}\leq
\|\boldsymbol{\Sigma}_{n,\rho}^{-1}\|_{L_{1}}\leq \rho^{-1} p$. The expectation
$\ep\|\hat{\boldsymbol{\Omega}}_{\rho}-\boldsymbol{\Omega}_{0}\|^{2}_{2}$ is then well-defined.
The other motivation to replace $\boldsymbol{\Sigma}_{n}$ with $\boldsymbol{\Sigma}_{n,\rho}$
comes from our implementation, which computes (\ref{c1}) by the primal
dual interior point method. One usually needs to specify a feasible
initialization. When $p>n$, it is hard to find an initial value for
(\ref{c1}). For (\ref{re-c1}), we can simply set the initial value to
$\boldsymbol{\Sigma}_{n,\rho}^{-1}$.
\begin{theorem}\label{thn-3} Suppose that
  $\boldsymbol{\Omega}_{0}\in\mathcal{U}(q,s_{0}(p))$ and (C1) holds. Let
  $\lambda_{n}=C_{0}M\sqrt{\log p/n}$ with $C_{0}$ being defined in
  Theorem \ref{thn-2} (i) and $\tau$ being sufficiently large. Let
  $\rho=\sqrt{\log p/n}.$ If $p\geq n^{\xi}$ for some $\xi>0$, then we
  have
  \begin{eqnarray*}
    \sup_{\boldsymbol{\Omega}_{0}\in
      \mathcal{U}}\ep\|\hat{\boldsymbol{\Omega}}_{\rho}-\boldsymbol{\Omega}_{0}\|^{2}_{2}=O\Big{(}
    M^{4-4q}s^{2}_{0}(p)\Big{(}\frac{\log p}{n}\Big{)}^{1-q}\Big{)}.
  \end{eqnarray*}
\end{theorem}

\medskip\noindent
{\bf Remark:} It is not necessary to restrict $\rho=\sqrt{\log
  p/n}$. In fact, from the proof we can see that Theorem \ref{thn-3}
still holds for
\begin{eqnarray}\label{a10}
  \min(\sqrt{\frac{\log p}{n}},p^{-\alpha})\leq \rho\leq\sqrt{\frac{\log
      p}{n}}
\end{eqnarray}
with any $\alpha>0$.

When the variables of $\textbf{X}$ are ordered, better rates can be
obtained. Similar as in Bickel and Levina (2008a), we
consider the following class of precision matrices:
\begin{eqnarray*}
  \mathcal{U}_{o}(\alpha,B)&=&\Big{\{}\boldsymbol{\Omega}:\boldsymbol{\Omega}\succ
  0,~~\max_{j}\sum_{i}\{|\omega_{ij}|:|i-j|\geq k\}\leq B
  (k+1)^{-\alpha}~\mbox{for all $k\geq 0$}\Big{\}}
\end{eqnarray*}
for $\alpha > 0$.  Suppose the modified Cholesky factor of
$\boldsymbol{\Omega}_{0}$ is $\boldsymbol{\Omega}_{0} = TD^{-1}T$, with the unique lower
triangular matrix $T$ and diagonal matrix $D$. To estimate
$\boldsymbol{\Omega}_{0}$, Bickel and Levina (2008a) used the banding method and
assumed $T\in \mathcal{U}_{o}(\alpha,B)$. It is easy to see that
$T\in \mathcal{U}_{o}(\alpha,B)$ implies $\boldsymbol{\Omega}_{0}\in
\mathcal{U}_{o}(\alpha,B_{1})$ for some constant $B_{1}$. Rather
than assuming $T\in \mathcal{U}_{o}(\alpha,B)$, we use a more
general assumption that $\boldsymbol{\Omega}_{0}\in \mathcal{U}_{o}(\alpha,B)$.

\begin{theorem}\label{th2} Let $\boldsymbol{\Omega}_{0}\in \mathcal{U}_{o}(\alpha,B)$ and
  $\lambda_{n}=CB\sqrt{\log p/n}$ with sufficiently large $C$.

  (i). If (C1) or (C2) holds, then with probability greater than
  $1-O\Big{(}n^{-\delta/8}+p^{-\tau/2}\Big{)}$,
  \begin{eqnarray}\label{c8}
    \|\hat{\boldsymbol{\Omega}}-\boldsymbol{\Omega}_{0}\|_{2}=O\Big{(}B^{2}\Big{(}\frac{\log
      p}{n}\Big{)}^{\alpha/(2\alpha+2)}\Big{)}.
  \end{eqnarray}

  (ii). Suppose that $p\geq n^{\xi}$ for some $\xi>0$. If (C1) holds
  and $\rho=\sqrt{\log p/n}$, then
  \begin{eqnarray}\label{c8-8}
    \sup_{\boldsymbol{\Omega}_{0}\in
      \mathcal{U}_{o}(\alpha,B)}\ep\|\hat{\boldsymbol{\Omega}}_{\rho}-\boldsymbol{\Omega}_{0}\|^{2}_{2}=
    O\Big{(}B^{4}\Big{(}\frac{\log
      p}{n}\Big{)}^{\alpha/(\alpha+1)}\Big{)}.
  \end{eqnarray}
\end{theorem}

Theorem \ref{th2} shows that our estimator has the same rate as that
in Bickel and Levina (2008a) by banding the Cholesky factor of the
precision matrix for the ordered variables.


\subsection{Rates under $l_{\infty}$ norm and  Frobenius norm}

We have so far focused on the performance of the estimator under the spectral norm loss.
Rates of convergence can also be obtained under the elementwise $l_{\infty}$ norm and
the Frobenius norm.

\begin{theorem}\label{cr1} (i). Under the conditions of Theorem \ref{thn-2} (i), we have
\begin{eqnarray*}
&&|\hat{\boldsymbol{\Omega}}-\boldsymbol{\Omega}_{0}|_{\infty}\leq 4C_{0}M^{2}\sqrt{\frac{\log
p}{n}},\cr &&\frac{1}{p}\|\hat{\boldsymbol{\Omega}}-\boldsymbol{\Omega}_{0}\|^{2}_{F}\leq
4C_{1}M^{4-2q}s_{0}(p)\Big{(}\frac{\log p}{n}\Big{)}^{1-q/2},
\end{eqnarray*}
with probability greater than $1-4p^{-\tau}$.

(ii).  Under the conditions of Theorem \ref{thn-2} (ii), we have
\begin{eqnarray*}
&&|\hat{\boldsymbol{\Omega}}-\boldsymbol{\Omega}_{0}|_{\infty}\leq 4C_{2}M^{2}\sqrt{\frac{\log
p}{n}},\cr &&\frac{1}{p}\|\hat{\boldsymbol{\Omega}}-\boldsymbol{\Omega}_{0}\|^{2}_{F}\leq
4C_{3}M^{4-2q}s_{0}(p)\Big{(}\frac{\log p}{n}\Big{)}^{1-q/2},
\end{eqnarray*}
with probability greater than
$1-O\Big{(}n^{-\delta/8}+p^{-\tau/2}\Big{)}$.
\end{theorem}

The rate in Theorem \ref{cr1} (ii) is significantly faster than the
one obtained by Ravikumar et al.\ (2008); see Section 3.3 for more
detailed discussions.   A similar rate to ours was obtained by Lam
and Fan (2009) under the Frobenius norm.  The elementwise $\ell_\infty$
norm result will lead to the model selection consistency result to
be shown in the next section.  We now give the rates for
$\hat{\boldsymbol{\Omega}}_{\rho}-\boldsymbol{\Omega}_{0}$ under expectation.

\begin{theorem}\label{n-cr1}  Under the conditions of Theorem \ref{thn-3}, we have
\begin{eqnarray*}
&&\sup_{\boldsymbol{\Omega}_{0}\in
\mathcal{U}}\ep|\hat{\boldsymbol{\Omega}}_{\rho}-\boldsymbol{\Omega}_{0}|^{2}_{\infty}=
O\Big{(}M^{4}\frac{\log p}{n}\Big{)},\cr
&&\frac{1}{p}\sup_{\boldsymbol{\Omega}_{0}\in
\mathcal{U}}\ep\|\hat{\boldsymbol{\Omega}}_{\rho}-\boldsymbol{\Omega}_{0}\|^{2}_{F}=
O\Big{(}M^{4-2q}s_{0}(p)\Big{(}\frac{\log
p}{n}\Big{)}^{1-q/2}\Big{)}.
\end{eqnarray*}
\end{theorem}


The proofs of  Theorems 1-5 rely on the following  more general
theorem.

\begin{theorem}\label{th1} Suppose that $\boldsymbol{\Omega}_{0}\in\mathcal{U}(q,s_{0}(p))$ and $\rho\geq 0$. If $\lambda_{n}\geq \|\boldsymbol{\Omega}_{0}\|_{L_{1}}
(\max_{ij}|\hat{\sigma}_{ij}-\sigma^{0}_{ij}|+\rho)$, then we have
\begin{eqnarray}\label{t2}
|\hat{\boldsymbol{\Omega}}_{\rho}-\boldsymbol{\Omega}_{0}|_{\infty}\leq
4\|\boldsymbol{\Omega}_{0}\|_{L_{1}}\lambda_{n},
\end{eqnarray}
\begin{eqnarray}\label{t1}
\|\hat{\boldsymbol{\Omega}}_{\rho}-\boldsymbol{\Omega}_{0}\|_{2}\leq C_{4}
s_{0}(p)\lambda_{n}^{1-q},
\end{eqnarray}
and
\begin{eqnarray}\label{t3}
\frac{1}{p}\|\hat{\boldsymbol{\Omega}}_{\rho}-\boldsymbol{\Omega}_{0}\|^{2}_{F}\leq
C_{5}s_{0}(p)\lambda_{n}^{2-q}
\end{eqnarray}
 where $C_{4}\leq
2(1+2^{1-q}+3^{1-q})(4\|\boldsymbol{\Omega}_{0}\|_{L_{1}})^{1-q}$ and $C_{5}\leq
4\|\boldsymbol{\Omega}_{0}\|_{L_{1}}C_{4}$.
\end{theorem}


\subsection{Comparison with lasso-type estimator}

We compare our results to those of Ravikumar et al.\ (2008), wherein
the authors estimated $\boldsymbol{\Omega}_{0}$ by solving the following $\ell_{1}$
regularized log-determinant program:
\begin{eqnarray}\label{lass}
  \hat{\boldsymbol{\Omega}}_{\star}:=\argmin_{\Theta\succ
    0}\{\langle\boldsymbol{\Omega},\boldsymbol{\Sigma}_{n}\rangle-\log\det(\boldsymbol{\Omega})+\lambda_{n}\|\boldsymbol{\Omega}\|_{1,\rm
    off}\},
\end{eqnarray}
where $\|\boldsymbol{\Omega}\|_{1,\rm off}=\sum_{i\neq j}|\omega_{ij}|$.  To obtain
the rates of convergence in the elementwise $\ell_{\infty}$ norm and
the spectral norm, they imposed the following condition:

\medskip\noindent {\bf Irrepresentable Condition in Ravikumar et al.\
  (2008)} There exists some $\alpha\in (0,1]$ such that
\begin{eqnarray}\label{ravicond}
  \|\boldsymbol{\Gamma}_{S^{c}S}(\boldsymbol{\Gamma}_{SS})^{-1}\|_{L_{1}}\leq 1-\alpha,
\end{eqnarray}
where $\boldsymbol{\Gamma}=\boldsymbol{\Sigma}^{-1}_{0}\otimes \boldsymbol{\Sigma}^{-1}_{0}$, $S$ is the
support of $\boldsymbol{\Omega}_{0}$ and $S^{c}=\{1, \dotsc, p\}\times
\{1,\dotsc,p\}-S$.

The above assumption is particularly strong. Under this assumption, it
was shown in Ravikumar et al.\ (2008) that $\hat{\boldsymbol{\Omega}}_{\star}$
estimates the zero elements of $\boldsymbol{\Omega}_{0}$ exactly by zero with high
probability. In fact, a similar condition to \eqref{ravicond} for
Lasso with the covariance matrix $\boldsymbol{\Sigma}_{0}$ taking the place of the
matrix $\boldsymbol{\Gamma}$ is sufficient and nearly necessary for recovering the
support using the ordinary Lasso; see for example Meinshausen and
B\"{u}hlmann (2006).

Suppose that $\boldsymbol{\Omega}_{0}$ is $s_{0}(p)$-sparse and consider
subgaussian random variables $X_{i}/\sqrt{\sigma^{0}_{ii}}$ with the
parameter $\sigma$. In addition to \eqref{ravicond}, Ravikumar et al.\
(2008) assumed that the sample size $n$ satisfies the bound
\begin{eqnarray}\label{c7}
  n>C_{1}s^{2}_{0}(p)(1+8/\alpha)^{2}(\tau\log p+\log 4),
\end{eqnarray}
where
$C_{1}=\{48\sqrt{2}(1+4\sigma^{2})\max_{i}(\sigma^{0}_{ii})\max\{\|\boldsymbol{\Sigma}_{0}\|_{L_{1}}K_{\boldsymbol{\Gamma}},\|\boldsymbol{\Sigma}_{0}\|^{3}_{L_{1}}K^{2}_{\boldsymbol{\Gamma}}\}\}^{2}$.
Under the aforementioned conditions, they showed that with probability
greater than $1-1/p^{\tau-2}$,
\begin{eqnarray*}
  |\hat{\boldsymbol{\Omega}}_{\star}-\boldsymbol{\Omega}_{0}|_{\infty}\leq
  \{16\sqrt{2}(1+4\sigma^{2})\max_{i}(\sigma_{ii})(1+8\alpha^{-1})K_{\boldsymbol{\Gamma}}\}\sqrt{\frac{\tau\log
      p+\log 4}{n}},
\end{eqnarray*}
where $K_{\boldsymbol{\Gamma}}=\|([\boldsymbol{\Sigma}_{0}\otimes
\boldsymbol{\Sigma}_{0}]_{SS})^{-1}\|_{L_{1}}$. Note that their constant depends on
quantities $\alpha$ and $K_{\boldsymbol{\Gamma}}$, while our constant depends on
$M$, the bound of $\|\boldsymbol{\Omega}_{0}\|_{L_{1}}$. They required (\ref{c7}),
while we only need $\log p=o(n)$. Another substantial difference is
that the irrepresentable condition (\ref{ravicond}) is not needed for
our results.

We next compare our result to that of Ravikumar et al.\ (2008) under
the case of polynomial-type tails. Suppose (C2) holds. Corollary 2 in
Ravikumar et al.\ (2008) shows that if
$p=O\Big{(}\{n/s^{2}_{0}(p)\}^{(\gamma+1+\delta/4)/\tau}\Big{)}$ for
some $\tau>2$, then with probability greater than $1-1/p^{\tau-2}$,
\begin{eqnarray*}
  |\hat{\boldsymbol{\Omega}}_{\star}-\boldsymbol{\Omega}_{0}|_{\infty}=O\Big{(}\sqrt{\frac{p^{\tau/(\gamma+1+\delta/4)}}{n}}\Big{)}.
\end{eqnarray*}
Theorem \ref{cr1} shows our estimator still enjoys the order of
$\sqrt{\log p/n}$ in the case of polynomial-type tails. Moreover, when
$\gamma\geq 1$, the range $p=O(n^{\gamma})$ in our theorem is wider
than their range
$p=O\Big{(}\{n/s^{2}_{0}(p)\}^{(\gamma+1+\delta/4)/\tau}\Big{)}$ with
$\tau>2$.

It is worth noting that instead of the sparse precision matrices, our
estimator allows for a wider class of matrices. For example, the
estimator is still consistent for the model which is not truly sparse
but has many small entries.

\section{Graphical Model Selection Consistency}
\label{sec:consistency}

As mentioned in the introduction, graphical model selection is an
important problem. The constrained $\ell_1$ minimization procedure
introduced in Section \ref{sec:estimation} for estimating $\boldsymbol{\Omega}_{0}$
can be modified to recover the support of $\boldsymbol{\Omega}_{0}$.  We introduce
an additional thresholding step based on $\hat{\boldsymbol{\Omega}}$. More
specifically, define a threshold estimator
$\tilde{\boldsymbol{\Omega}}=(\tilde{\omega}_{ij})$ with
\begin{eqnarray*}
  \tilde{\omega}_{ij}=\hat{\omega}_{ij}I\{|\hat{\omega}_{ij}|\geq
  \tau_{n}\},
\end{eqnarray*}
where $\tau_{n}\geq 4M\lambda_{n}$ is a tuning parameter and
$\lambda_{n}$ is given in Theorem \ref{thn-2}.

Define
\begin{eqnarray*}
  &&\mathcal{M}(\tilde{\boldsymbol{\Omega}})=\{\sgn(\tilde{\omega}_{ij}),~~1\leq
  i,j\leq p\},\cr
  &&\mathcal{M}(\boldsymbol{\Omega}_{0})=\{\sgn(\omega^{0}_{ij}),~~1\leq i,j\leq
  p\},\cr
  &&S(\boldsymbol{\Omega}_{0})=\{(i,j): \omega^{0}_{ij}\neq 0\},
\end{eqnarray*}
and
\begin{eqnarray*}
  \theta_{\min}=\min_{(i,j)\in S(\boldsymbol{\Omega}_{0})}|\omega^{0}_{ij}|.
\end{eqnarray*}
From the elementwise $\ell_\infty$ results established in Theorem \ref{cr1},
with high probability, the resulting elements in $\hat{\boldsymbol{\Omega}}$ shall
exceed the threshold level if the corresponding element in $\boldsymbol{\Omega}_0$
is large in magnitude.  On the contrary, the elements of
$\hat{\boldsymbol{\Omega}}$ outside the support of $\boldsymbol{\Omega}_0$ will remain below the
threshold level with high probability.  Therefore, we have the
following theorem on the threshold estimator $\tilde{\boldsymbol{\Omega}}$.

\begin{theorem} \label{thm:supp} Suppose that (C1) or (C2) holds and $\boldsymbol{\Omega}_0 \in
  \mathcal{U}(0,s_{0}(p))$. If $\theta_{\min}>2 \tau_{n}$, then with probability greater than
$1-O\Big{(}n^{-\delta/8}+p^{-\tau/2}\Big{)}$, we have
  $\mathcal{M}(\tilde{\boldsymbol{\Omega}})=\mathcal{M}(\boldsymbol{\Omega}_{0})$.
\end{theorem}

The threshold estimator $\tilde{\boldsymbol{\Omega}}$ not only recovers the
sparsity pattern of $\boldsymbol{\Omega}_{0}$, but also recovers the signs of the
nonzero elements. This property is called sign consistency in some
literature.

The condition $\theta_{\min}>2 \tau_{n}$ is needed to ensure that
nonzero elements are correctly retained. From Theorem \ref{cr1}, we
see that, if $M$ does not depend on $n,p$, then $\tau_{n}$ is of order
$\sqrt{\log p/n}$ which is the same order as in Ravikumar et al.\
(2008) for exponential-type tails, but weaker than their assumption
$\theta_{\min}\geq C\sqrt{\frac{p^{\tau/(\gamma+1+\delta/4)}}{n}}$ for
polynomial-type tails.

Based on Meinshausen and B\"{u}hlmann (2006), Zhou et al.\ (2009)
applied adaptive Lasso to covariance selection in Gaussian graphical
models.  For $\textbf{X}=(X_{1},\dotsc,X_{p})\sim N(\boldsymbol{0},\boldsymbol{\Sigma}_{0})$,
they regress $X_{i}$ versus the other variables $\{X_{k}; k\neq i\}$:
$X_{i}=\sum_{j\neq i}\beta^{i}_{j}X_{j}+V_{i}$, where $V_{i}$ is a
normally distributed random variables with mean zero and the
underlying coefficients can be shown to be
$\beta^{i}_{j}=-\omega^{0}_{ij}/\omega^{0}_{ii}$. Then they use the
adaptive Lasso to recover the support of $\{\beta^{i}_{j}\}$, which is
identical to the support of $\boldsymbol{\Omega}_{0}$. A main assumption in their
paper is the restricted eigenvalue assumption on $\boldsymbol{\Sigma}_{0}$ which is
weaker than the irrepresentable condition. Their method can recover
the support of $\boldsymbol{\Omega}_{0}$ but is unable to estimate the elements in
$\boldsymbol{\Omega}_{0}$. Without imposing the unnecessary irrepresentable
condition, the additional advantage of our method is that it not only
recovers the support of $\boldsymbol{\Omega}_{0}$ but also provides consistency results
under the elementwise $l_{\infty}$ norm and the spectral norm.

\section{Numerical Results}
\label{sec:simu}

In this section we turn to the numerical performance of our CLIME
estimator. The procedure is easy to implement.  An R package of our
method has been developed and is available on the web at\\
\verb+http://stat.wharton.upenn.edu/~tcai/paper/html/Precision-Matrix.html+.\\
The goal of this section is to first investigate the numerical
performance of the estimator through simulation studies and then apply
our method to the analysis of a breast cancer dataset.

The proposed estimator $\hat{\boldsymbol{\Omega}}$ can be obtained in a column by
column fashion as illustrated in Lemma 1.  Hence we will focus on the
numerical implementation of solutions to the optimization problem
\eqref{o1}:
\begin{equation*}
  \min|\boldsymbol{\beta}|_{1} ~~\mbox{subject to}~~ |\boldsymbol{\Sigma}_{n}\boldsymbol{\beta} - \boldsymbol{e}_i|_{\infty}\leq
  \lambda_{n}.
\end{equation*}
We consider  relaxation of the above, which is equivalent to the
following linear programming problem:
\begin{equation}
  \begin{split}\label{eq:lin}
    &\min  \sum_{j=1}^p u_j \\
    \text{subject to: } & - \beta_j \le u_j \text{ for all }1\le j \le
    p\\
    & + \beta_j \le u_j \text{ for all }1 \le j \le
    p\\
    & -\hat{\sigma}_k^T \boldsymbol{\beta} + I\{k = i\} \le \lambda_n \text{ for all }1\le
    k \le
    p \\
    & + \hat{\sigma}_k^T \boldsymbol{\beta} - I\{k = i\} \le \lambda_n \text{ for all
    }1\le k \le p.
  \end{split}
\end{equation}
The same linear relaxation was considered in Cand\`{e}s and Tao
(2007), and was shown there to be very efficient for the Dantzig
selector problem in regression. To solve \eqref{eq:lin}, we follow the
primal dual interior method approach, for example see Boyd and
Vandenberghe (2004).  The resulting algorithm has comparable numerical
performance as other numerical procedures, for example Glasso.  Note
that we only need sweep through the $p$ columns once but Glasso does
need to have an extra outer layer of iterations to loop through the
$p$ columns several times by cyclical coordinate descent. Once
$\hat{\boldsymbol{\Omega}}_1$ is obtained by combining the $\hat{\boldsymbol{\beta}}$'s for each
column, we symmetrize $\hat{\boldsymbol{\Omega}}_1$ by setting the entry $(i,j)$ to
be the smaller one in magnitude of two entries $\hat{\omega}_{ij}^1$
and $\hat{\omega}_{ji}^1$, for all $1\le i, j \le p$, as in
\eqref{me1}.

Similar to many iterative methods, our method also requires a proper
initialization within the feasible set.  The initializing $\boldsymbol{\beta}^0$
however cannot be simply replaced by the solution of the linear system
$\boldsymbol{\Sigma}_n \boldsymbol{\beta} = \boldsymbol{e}_i$ for each $i$ when $p> n$, since $\boldsymbol{\Sigma}_n$ is
singular.  The remedy is to add a small positive constant $\rho$
(e.g. $\rho = \sqrt{\log p /n}$) to all the diagonal entries of the
matrix $\boldsymbol{\Sigma}_n$, that is we use the $\rho$-perturbed matrix
$\boldsymbol{\Sigma}_{n,\rho} = \boldsymbol{\Sigma}_n + \rho \boldsymbol{I}$ to replace the $\boldsymbol{\Sigma}_n$ in
\eqref{eq:lin}.  Such a perturbation does not noticeably affect the
computational accuracy of the final solution in our numerical
experiments.  The resulting solution $\hat{\boldsymbol{\Omega}}_{\rho}$ in the
perturbed problem \eqref{re-c1} is shown to have all the theoretical
properties in Sections 3 and 4, and even better the convergence rate
of the spectral norm under expectation is also established there for
$\hat{\boldsymbol{\Omega}}_{\rho}$.

In the context of high dimensional linear regression, a second stage
refitting procedure was considered in Cand\'{e}s and Tao (2007) to
correct the biases introduced by the $\ell_1$ norm penalization.
Their refitting procedure seeks the best coefficient vector, giving
the maximum likelihood, which has  the
same support as the original Dantzig selector.  Inspired by this two-stage procedure, we propose a
similar two-stage procedure to further improve the numerical
performance of the CLIME estimator by refitting as
\begin{equation*}
  \check{\boldsymbol{\Omega}} = \argmin_{\boldsymbol{\Omega}_{\hat{S}^c}  = 0}
  \{\langle\boldsymbol{\Omega},\boldsymbol{\Sigma}_{n}\rangle-\log\det(\boldsymbol{\Omega})\}
\end{equation*}
where $\hat{S} = S(\tilde{\boldsymbol{\Omega}})$ and $\boldsymbol{\Omega}_{\hat{S}^c} =
\{\omega_{ij},\; (i,j) \in \hat{S}^c \}$.  Here the estimator
$\check{\boldsymbol{\Omega}}$ minimizes the Bregman divergence among all symmetric
positive definite matrices under the constraint.  We shall call
$\check{\boldsymbol{\Omega}}$ Refitted CLIME hereafter.
The bounds under the three norms in Section~\ref{sec:rate} and the
support recovery $S(\check{\boldsymbol{\Omega}}) = S(\boldsymbol{\Omega}_0)$ can also be
established. For example, the Frobenius loss bound can be easily
derived from the same approach used in Rothman et al.\ (2008) and Fan
et al.\
(2009).  
Other theoretical properties are more involved and we leave this to
future work.  

\subsection{Simulations}

We now compare the numerical performance of the CLIME estimator
$\hat{\boldsymbol{\Omega}}_{\rm CLIME}$, the Refitted CLIME estimator, the
Graphical Lasso $\hat{\boldsymbol{\Omega}}_{\rm Glasso}$ and the SCAD
$\hat{\boldsymbol{\Omega}}_{\rm SCAD}$ from Fan et al.\ (2009) which is defined as
\begin{equation*}
  \hat{\boldsymbol{\Omega}}_{\rm SCAD} := \argmin_{\Theta\succ
    0}\{\langle\boldsymbol{\Omega},\boldsymbol{\Sigma}_{n}\rangle-\log\det(\boldsymbol{\Omega})+ \sum_{i=1}^p
  \sum_{j=1}^p \textrm{SCAD}_{\lambda, a}(\lvert \omega_{ij} \rvert)
  \}.
\end{equation*}
where the SCAD function $\textrm{SCAD}_{\lambda, a}$ is proposed by
Fan (1997).  We use recommended choice $a=3.7$ by Fan and Li (2001)
throughout and set all $\lambda$ to be the same for all $(i,j)$
entries for simplicity.  This setting for $a$ and $\lambda$ is the
same as that of Fan et al.\ (2009). See Fan et al.\ (2009) for further
details on $\hat{\boldsymbol{\Omega}}_{\rm SCAD}$.
Note that $\hat{\boldsymbol{\Omega}}_{\rm Glasso}$ has the equivalent performance
as the SPICE estimator by Rothman et al.\ (2008) according to their
study.

We consider three models as follows:
\begin{itemize}
\item Model 1. $\omega^{0}_{ij}=0.6^{|i-j|}$.
\item Model 2. The second model comes from Rothman et al.\ (2008). We
  let $\boldsymbol{\Omega}_{0}=\boldsymbol{B}+\delta \boldsymbol{I}$, where each off-diagonal entry in $\boldsymbol{B}$ is
  generated independently and equals to 0.5 with probability 0.1 or 0
  with probability 0.9.  $\delta$ is chosen such that the conditional
  number (the ratio of maximal and minimal singular values of a
  matrix) is equal to $p$.  Finally, the matrix is standardized to
  have unit diagonals.
\item Model 3. In this model, we consider a non-sparse matrix and let
  $\boldsymbol{\Omega}_{0}$ have all off-diagonal elements $0.5$ and the diagonal
  elements $1$.
\end{itemize}
The first model has a banded structure, and the values of the entries
decay as they move away from the diagonal.  The second is an example
of a sparse matrix without any special sparsity patterns.  The third
serves as a dense matrix example.

For each model, we generate a training sample of size $n=100$ from a
multivariate normal distribution with mean zero and covariance matrix
$\boldsymbol{\Sigma}_0$, and an independent sample of size $100$ from the same
distribution for validating the tuning parameter $\lambda$.  Using the
training data, a series of estimators with $50$ different values of
$\lambda$ are computed, and the one with the smallest likelihood loss
on the validation sample is used, where the likelihood loss is defined
by
\begin{align*}
  L(\boldsymbol{\Sigma}, \boldsymbol{\Omega}) = \langle \boldsymbol{\Omega}, \boldsymbol{\Sigma} \rangle - \log
  \det(\boldsymbol{\Omega}).
\end{align*}
The Glasso and SCAD estimators are computed on the same training and
testing data using the same cross validation scheme. We consider
different values of $p=30,60,90,120, 200$ and replicate $100$ times.


The estimation quality is first measured by the following matrix
norms: the operator norm, the matrix $\ell_1$ norm and the Frobenius
norm.
Table \ref{tb:simu} reports the averages and standard errors of
these losses.
\begin{sidewaystable}[hp!]
  \begin{center}
    \caption{Comparison of average(SE) matrix losses for three models
      over $100$ replications.}
    \begin{tabular}{|c c c c @{\hspace{2em}} c c c @{\hspace{2em}} c c
        c|}
      \multicolumn{10}{c}{Operator norm}\\
      \hline
      & \multicolumn{3}{c}{Model 1} &\multicolumn{3}{c}{Model
        2}&\multicolumn{3}{c|}{Model 3} \\
      \hline
      $p$&  $\hat{\boldsymbol{\Omega}}_{\rm CLIME}$ &     $\hat{\boldsymbol{\Omega}}_{\rm
        Glasso}$  & $\hat{\boldsymbol{\Omega}}_{\rm SCAD}$ &
      $\hat{\boldsymbol{\Omega}}_{\rm CLIME}$ &     $\hat{\boldsymbol{\Omega}}_{\rm Glasso}$ & $\hat{\boldsymbol{\Omega}}_{\rm SCAD}$ &
      $\hat{\boldsymbol{\Omega}}_{\rm CLIME}$
      &     $\hat{\boldsymbol{\Omega}}_{\rm Glasso}$ &  $\hat{\boldsymbol{\Omega}}_{\rm SCAD}$
      \\
      30 & $2.28(0.02)$&$2.48(0.01)$&$2.38(0.02)$&$0.74(0.01)$&$0.77(0.01)$&$0.59(0.02)$&$14.95(0.004)$&$14.96(0.004)$&$14.97(0.002)$ \\
      60 & $2.79(0.01)$&$2.93(0.01)$&$2.71(0.01)$&$1.13(0.01)$&$1.12(0.01)$&$0.95(0.01)$&$30.01(0.002)$&$30.02(0.002)$&$29.98(0.001)$ \\
      90 & $2.97(0.01)$&$3.07(0.004)$&$2.76(0.004)$&$1.69(0.01)$&$1.49(0.004)$&$1.14(0.01)$&$45.01(0.002)$&$45.03(0.001)$&$44.98(0.001)$ \\
      120 & $3.08(0.004)$&$3.14(0.003)$&$2.79(0.004)$&$2.16(0.01)$&$1.82(0.003)$&$1.38(0.01)$&$60.01(0.002)$&$60.04(0.001)$&$58.40(0.10)$ \\
      200 & $3.17(0.01)$&$3.25(0.002)$&$2.83(0.003)$&$2.36(0.01)$&$2.46(0.002)$&$2.11(0.01)$&$100.02(0.001)$&$100.08(0.001)$&$96.69(0.01)$ \\
      \hline
      \multicolumn{10}{c}{}\\
      \multicolumn{10}{c}{Matrix $\ell_1$-norm}\\
      \hline
      & \multicolumn{3}{c}{Model 1} &\multicolumn{3}{c}{Model
        2}&\multicolumn{3}{c|}{Model 3} \\
      \hline
      $p$&  $\hat{\boldsymbol{\Omega}}_{\rm CLIME}$&     $\hat{\boldsymbol{\Omega}}_{\rm
        Glasso}$  & $\hat{\boldsymbol{\Omega}}_{\rm SCAD}$ &
      $\hat{\boldsymbol{\Omega}}_{\rm CLIME}$ &     $\hat{\boldsymbol{\Omega}}_{\rm Glasso}$ & $\hat{\boldsymbol{\Omega}}_{\rm SCAD}$ &
      $\hat{\boldsymbol{\Omega}}_{\rm CLIME}$
      &     $\hat{\boldsymbol{\Omega}}_{\rm Glasso}$  & $\hat{\boldsymbol{\Omega}}_{\rm SCAD}$  \\
      30 & $2.91(0.02)$&$3.08(0.01)$&$2.91(0.02)$&$1.29(0.02)$&$1.36(0.01)$&$0.81(0.02)$&$15.12(0.004)$&$15.08(0.003)$&$15.10(0.002)$ \\
      60 & $3.32(0.01)$&$3.55(0.01)$&$3.11(0.01)$&$2.10(0.02)$&$2.11(0.02)$&$1.98(0.03)$&$30.17(0.002)$&$30.15(0.002)$&$30.12(0.002)$ \\
      90 & $3.44(0.01)$&$3.72(0.01)$&$3.19(0.01)$&$2.95(0.02)$&$2.87(0.02)$&$2.71(0.03)$&$45.18(0.002)$&$45.18(0.002)$&$45.13(0.002)$ \\
      120 & $3.48(0.01)$&$3.81(0.01)$&$3.24(0.01)$&$3.69(0.02)$&$3.33(0.02)$&$3.32(0.03)$&$60.20(0.002)$&$60.20(0.003)$&$60.55(0.06)$ \\
      200 & $3.55(0.01)$&$4.01(0.01)$&$3.37(0.01)$&$4.13(0.02)$&$4.52(0.02)$&$4.67(0.03)$&$100.22(0.002)$&$100.24(0.002)$&$102.64(0.05)$ \\
      \hline
      \multicolumn{10}{c}{}\\
      \multicolumn{10}{c}{Frobenius norm}\\
      \hline
      & \multicolumn{3}{c}{Model 1} &\multicolumn{3}{c}{Model
        2}&\multicolumn{3}{c|}{Model 3} \\
      \hline
      $p$&  $\hat{\boldsymbol{\Omega}}_{\rm CLIME}$ &     $\hat{\boldsymbol{\Omega}}_{\rm
        Glasso}$  & $\hat{\boldsymbol{\Omega}}_{\rm SCAD}$ &
      $\hat{\boldsymbol{\Omega}}_{\rm CLIME}$ &     $\hat{\boldsymbol{\Omega}}_{\rm Glasso}$ & $\hat{\boldsymbol{\Omega}}_{\rm SCAD}$ &
      $\hat{\boldsymbol{\Omega}}_{\rm CLIME}$
      &     $\hat{\boldsymbol{\Omega}}_{\rm Glasso}$  & $\hat{\boldsymbol{\Omega}}_{\rm SCAD}$ \\
      30 & $3.81(0.04)$&$4.23(0.03)$&$3.97(0.03)$&$1.72(0.02)$&$1.71(0.01)$&$1.23(0.02)$&$14.96(0.004)$&$14.97(0.004)$&$14.97(0.001)$ \\
      60 & $6.63(0.03)$&$7.14(0.02)$&$6.37(0.02)$&$3.33(0.02)$&$3.10(0.01)$&$3.11(0.01)$&$30.02(0.002)$&$30.02(0.002)$&$29.98(0.001)$ \\
      90 & $8.78(0.04)$&$9.25(0.01)$&$7.98(0.01)$&$4.92(0.02)$&$4.36(0.01)$&$4.51(0.01)$&$45.02(0.002)$&$45.04(0.001)$&$44.99(0.001)$ \\
      120 & $10.58(0.02)$&$10.97(0.01)$&$9.31(0.01)$&$6.50(0.03)$&$5.50(0.01)$&$5.89(0.01)$&$60.01(0.001)$&$60.05(0.001)$&$60.60(0.08)$ \\
      200 & $14.20(0.04)$&$14.85(0.01)$&$12.21(0.01)$&$7.57(0.02)$&$8.15(0.01)$&$8.41(0.01)$&$100.02(0.001)$&$100.08(0.001)$&$103.41(0.02)$ \\
      \hline
    \end{tabular}
    \label{tb:simu}
  \end{center}
\end{sidewaystable}

We see that CLIME nearly uniformly outperforms Glasso.  The
improvement tends to be slightly more significant for sparse models
when $p$ is large, but overall the improvement is not dramatic.  Among
the three methods, SCAD is computationally most costly, but
numerically it has the best performance among the three when $p < n$
and is comparable to CLIME when $p$ is large.
Note that SCAD employs a nonconvex penalty to correct the bias while
CLIME currently optimizes the convex $\ell_1$ norm objective
efficiently.  
A more comparable procedure that also corrects the bias is our
two-stage Refitted CLIME, denoted by $\hat{\boldsymbol{\Omega}}_{\rm R-CLIME}$.
Table~\ref{tab:gclime} illustrates the improvement from bias
correction, and we only list the spectral norm loss for reasons of
space.  It is clear that our Refitted CLIME estimator has comparable
or better performance than SCAD, and our Refitted CLIME is especially
favorable when $p$ is large.
\begin{table}[htb!]
  \centering
  \caption{Comparison of average(SE)  operator norm  losses  from Model
    1  and 2 over $100$ replications.}
  \begin{tabular}[htb!]{| c c c c c |}
    \hline
    & \multicolumn{2}{c}{Model 1}  &  \multicolumn{2}{c|}{Model 2}  \\
    \hline
    $p$ & $\hat{\boldsymbol{\Omega}}_{\rm R-CLIME}$ & $\hat{\boldsymbol{\Omega}}_{\rm SCAD}$ & $\hat{\boldsymbol{\Omega}}_{\rm R-CLIME}$
    & $\hat{\boldsymbol{\Omega}}_{\rm SCAD} $\\
    30 & $1.56(0.02)$  &  $2.38(0.02)$ &  $0.85(0.01)$    &   $0.59(0.02)$  \\
    60 &     $2.15(0.01)$ & $2.71(0.01)$ &   $1.14(0.01)$  &  $0.95(0.09)$  \\
    90   & $2.42(0.01)$  & $2.76(0.004)$ & $1.17(0.01)$     &   $1.14(0.01)$ \\ 
    120  & $2.56(0.01)$  & $2.79(0.004)$ &  $1.44(0.01)$    &   $1.38(0.01)$ \\ 
    200 & $2.71(0.01)$ & $2.83(0.003)$ &  $1.91(0.01)$     &   $2.11(0.01)$  \\
    \hline
  \end{tabular}
  \label{tab:gclime}
\end{table}

Gaussian graphical model selection has also received considerable
attention in the literature. As we discussed earlier, this is
equivalent to the support recovery of the precision matrix. The
proportion of true zero (TN) and nonzero (TP) elements recovered by
two methods are also reported here in Table \ref{tb:simu2}.  The
numerical values over $10^{-3}$ in magnitude are considered to be
nonzero since the computation accuracy is set to be $10^{-4}$.
\begin{sidewaystable}[hp!]
  \begin{center}
    \caption{Comparison of average(SE) support recovery for three models over
      $100$ replications.}
    \begin{tabular}{|c c c c @{\hspace{2em}} c c c @{\hspace{2em}} c c
        c|}
      \multicolumn{10}{c}{TN\%}\\
      \hline
      & \multicolumn{3}{c}{Model 1} &\multicolumn{3}{c}{Model
        2}&\multicolumn{3}{c|}{Model 3} \\
      \hline
      $p$&  $\hat{\boldsymbol{\Omega}}_{\rm CLIME}$&     $\hat{\boldsymbol{\Omega}}_{\rm
        Glasso}$  &  $\hat{\boldsymbol{\Omega}}_{\rm SCAD}$  &
      $\hat{\boldsymbol{\Omega}}_{\rm CLIME}$ &     $\hat{\boldsymbol{\Omega}}_{\rm Glasso}$ &  $\hat{\boldsymbol{\Omega}}_{\rm SCAD}$ &
      $\hat{\boldsymbol{\Omega}}_{\rm CLIME}$
      &     $\hat{\boldsymbol{\Omega}}_{\rm Glasso}$ &     $\hat{\boldsymbol{\Omega}}_{\rm SCAD}$  \\
      \hline
      30 &
      $78.69(0.61)$&$50.65(0.75)$&$99.26(0.17)$&$77.41(0.86)$&$64.70(0.42)$&$99.10(0.08)$&
      N/A & N/A & N/A \\
      60 &
      $90.37(0.27)$&$69.47(0.29)$&$99.86(0.03)$&$85.98(0.36)$&$69.44(0.21)$&$96.08(0.14)$&
      N/A & N/A & N/A \\
      90 &
      $94.30(0.27)$&$77.62(0.20)$&$99.88(0.02)$&$91.15(0.17)$&$71.57(0.15)$&$95.98(0.11)$&
      N/A & N/A & N/A \\
      120 &
      $96.45(0.06)$&$81.46(0.16)$&$99.91(0.01)$&$94.87(0.19)$&$75.33(0.10)$&$95.69(0.10)$&
      N/A & N/A & N/A \\
      200 &
      $97.41(0.11)$&$85.36(0.11)$&$99.92(0.01)$&$81.74(0.26)$&$66.07(0.12)$&$96.97(0.05)$&
      N/A & N/A & N/A \\
      \hline
      \multicolumn{10}{c}{}\\
      \multicolumn{10}{c}{TP\%}\\
      \hline
      & \multicolumn{3}{c}{Model 1} &\multicolumn{3}{c}{Model
        2}&\multicolumn{3}{c|}{Model 3} \\
      \hline
      $p$&  $\hat{\boldsymbol{\Omega}}_{\rm CLIME}$&     $\hat{\boldsymbol{\Omega}}_{\rm
        Glasso}$  & $\hat{\boldsymbol{\Omega}}_{\rm SCAD}$  &
      $\hat{\boldsymbol{\Omega}}_{\rm CLIME}$ &     $\hat{\boldsymbol{\Omega}}_{\rm Glasso}$ &  $\hat{\boldsymbol{\Omega}}_{\rm SCAD}$ &
      $\hat{\boldsymbol{\Omega}}_{\rm CLIME}$
      &     $\hat{\boldsymbol{\Omega}}_{\rm Glasso}$ &     $\hat{\boldsymbol{\Omega}}_{\rm SCAD}$  \\
      \hline
      30 & $41.07(0.58)$&$60.20(0.56)$&$16.93(0.28)$&$99.66(0.09)$&$99.98(0.02)$&$97.70(0.24)$&$14.88(0.50)$&$20.07(0.57)$&$3.38(0.001)$ \\
      60 & $25.96(0.30)$&$41.72(0.32)$&$12.72(0.15)$&$85.10(0.36)$&$96.47(0.13)$&$79.81(0.44)$&$6.86(0.05)$&$10.49(0.20)$&$1.67(0.001)$ \\
      90 & $20.32(0.32)$&$33.70(0.23)$&$11.94(0.09)$&$66.25(0.39)$&$91.62(0.15)$&$67.93(0.48)$&$5.86(0.03)$&$7.54(0.13)$&$1.11(0.001)$ \\
      120 & $17.16(0.09)$&$29.32(0.20)$&$11.57(0.07)$&$42.37(0.49)$&$82.45(0.15)$&$54.92(0.41)$&$5.11(0.02)$&$6.20(0.12)$&$20.63(2.47)$ \\
      200 & $15.03(0.13)$&$25.34(0.15)$&$11.07(0.06)$&$57.07(0.27)$&$73.43(0.14)$&$30.50(0.40)$&$3.56(0.01)$&$4.94(0.02)$&$39.76(0.02)$ \\
      \hline
    \end{tabular}
    \label{tb:simu2}
  \end{center}
\end{sidewaystable}

\begin{figure}[h!]
  \centering \subfloat[][Model
  1]{\includegraphics[width=0.4\textwidth]{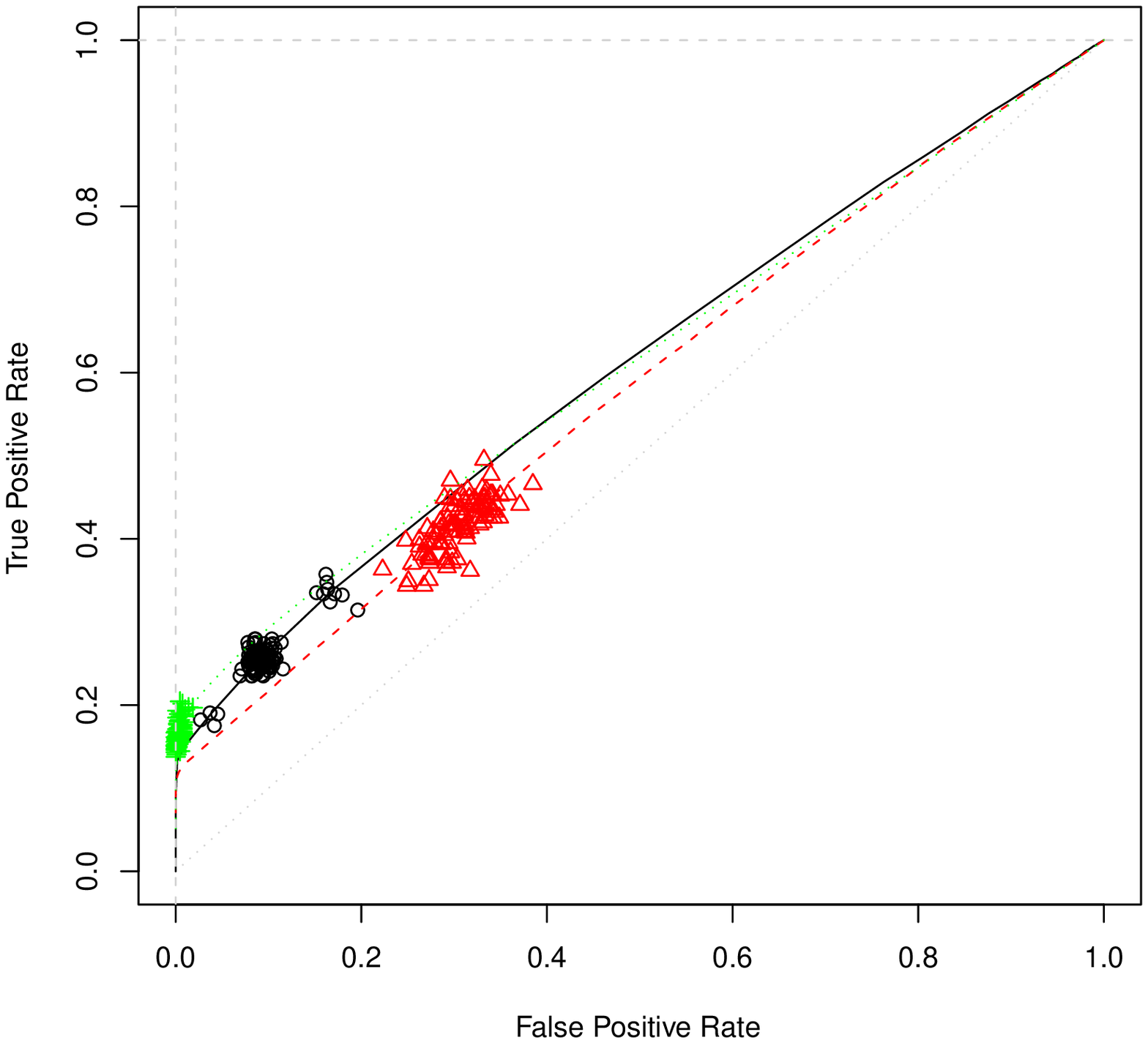}}
  \hspace*{0.06\textwidth} \subfloat[][Model
  2]{\includegraphics[width=0.4\textwidth]{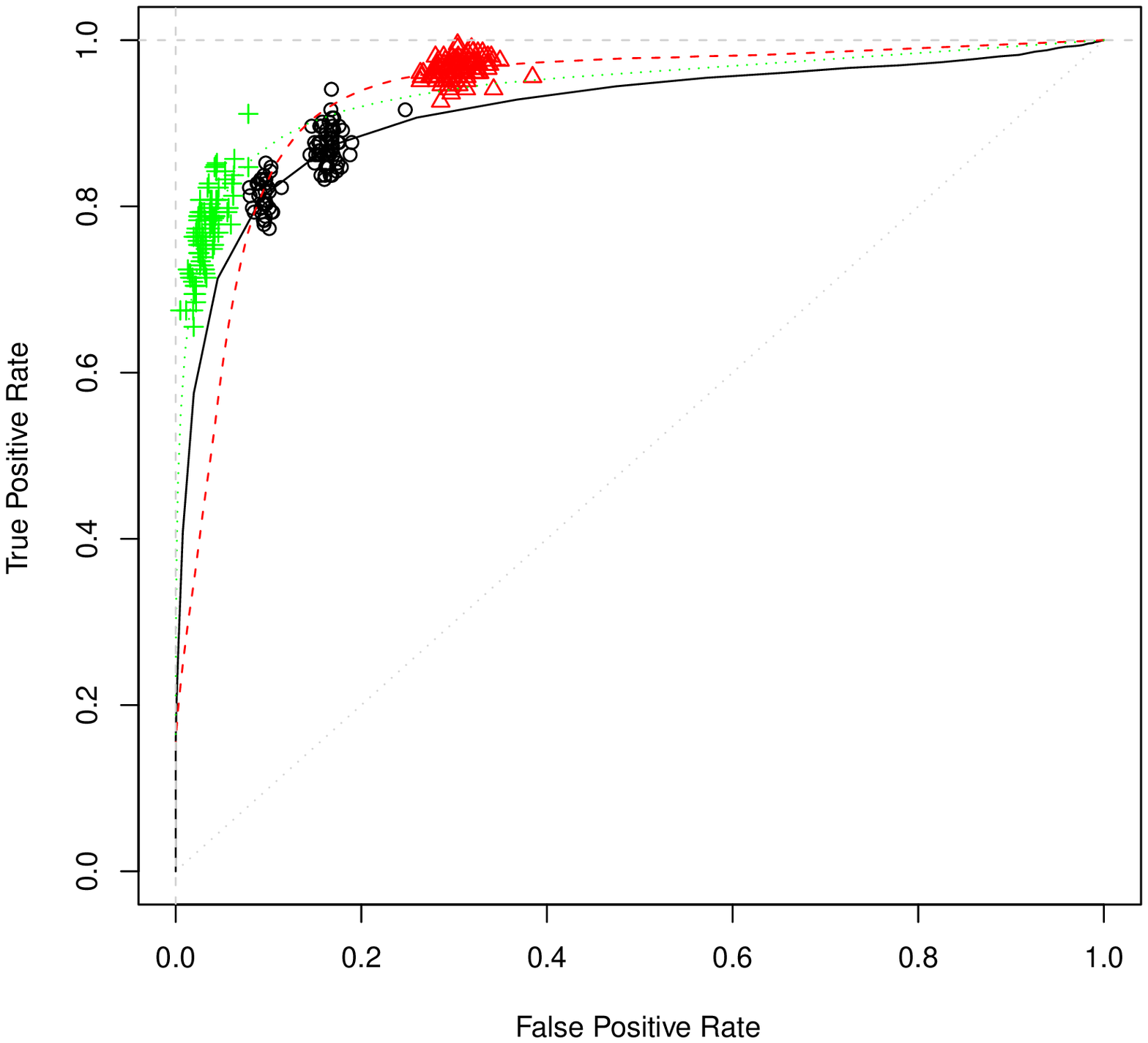}}
  \caption{TPR vs FPR for $p=60$.  The 
    solid, 
    dashed and
    dotted lines are the average TPR and FPR values for CLIME, Glasso
    and SCAD respectively as the tuning parameters of these methods
    vary.  The 
    circles, 
    triangles and 
    pluses correspond to $100$ different realizations of CLIME, Glasso
    and SCAD respectively, with the tuning parameter picked by cross
    validation.}
  \label{fig:roc}
\end{figure}

It is noticeable that Glasso tends to be more noisy by including
erroneous nonzero elements; CLIME tends to be more sparse than Glasso,
which is usually favorable in real applications; SCAD produces the
most sparse among the three but with a price of erroneously estimating
more true nonzero entries by zero.  This conclusion can also be
reached in Figure~\ref{fig:roc}, where the TPR and FPR values of $100$
realizations of these three procedures for first two models (note that
all elements in Model $3$ are nonzero) are plotted for $p=60$ as a
representative example of other cases.

\begin{figure}[h!]
  \centering
  Model 1\\
  \subfloat[][Truth
  ]{\includegraphics[width=0.19\textwidth]{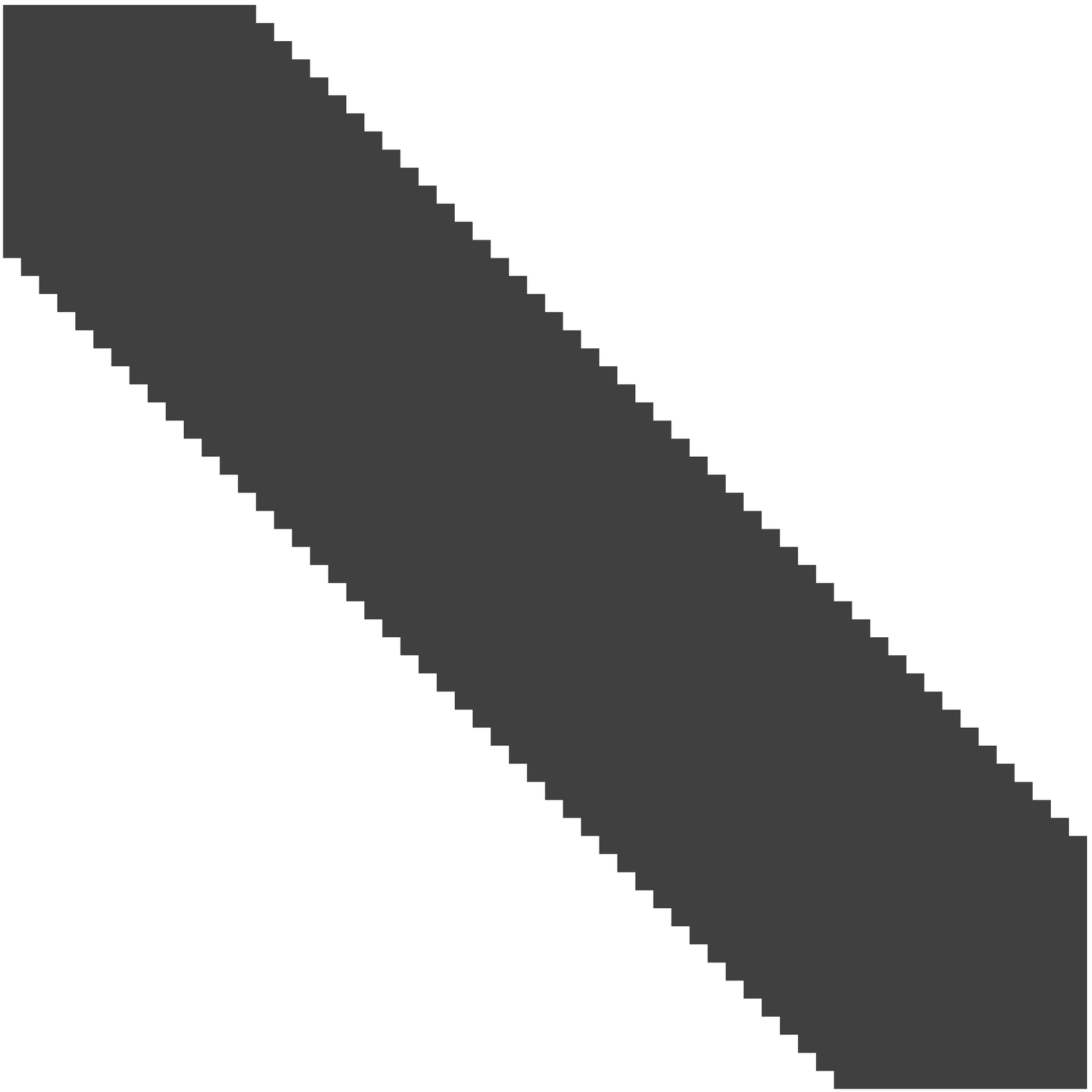} }
  \hspace*{0.06\textwidth} \subfloat[][CLIME
  ]{\includegraphics[width=0.19\textwidth]{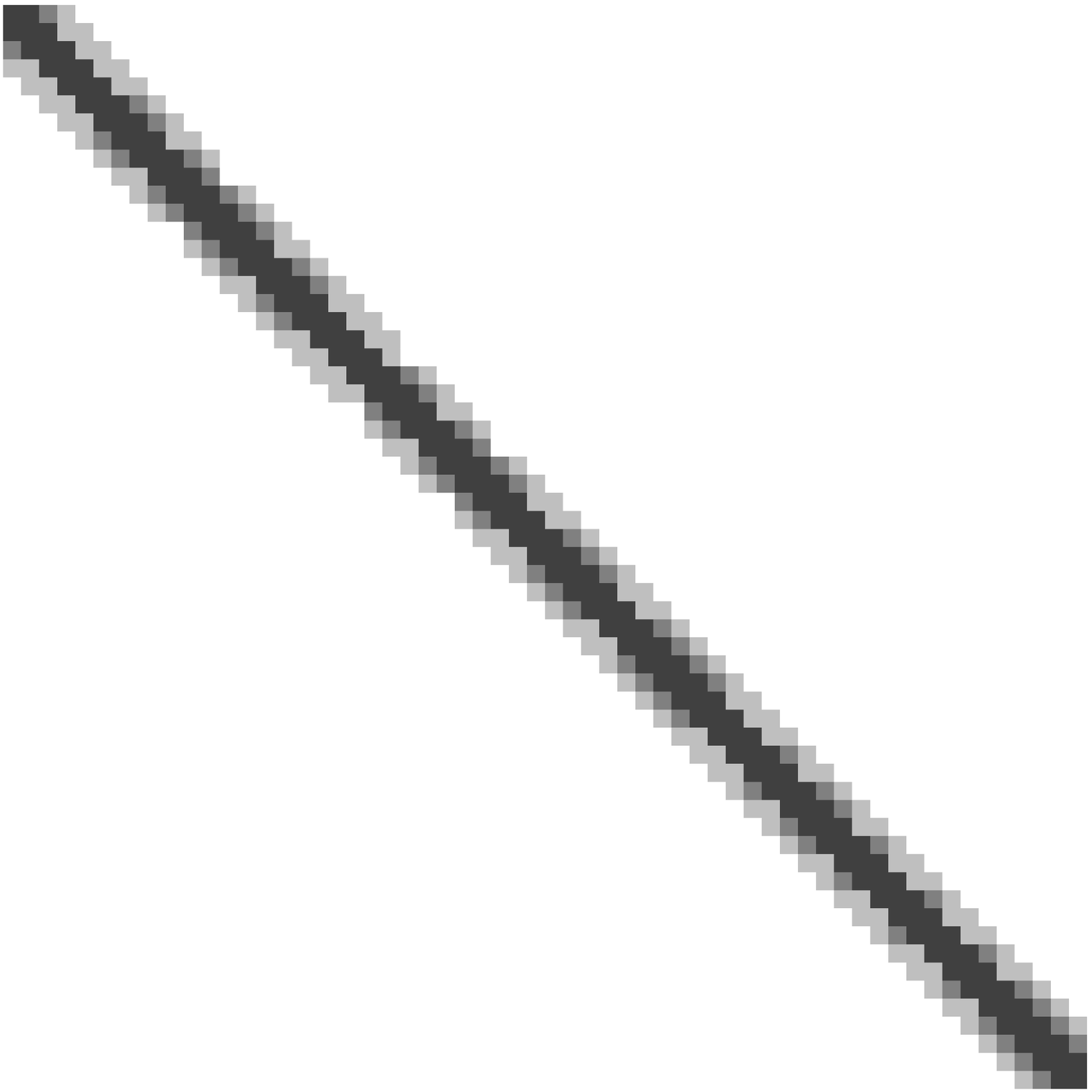} }
  \hspace*{0.06\textwidth} \subfloat[][Glasso
  ]{\includegraphics[width=0.19\textwidth]{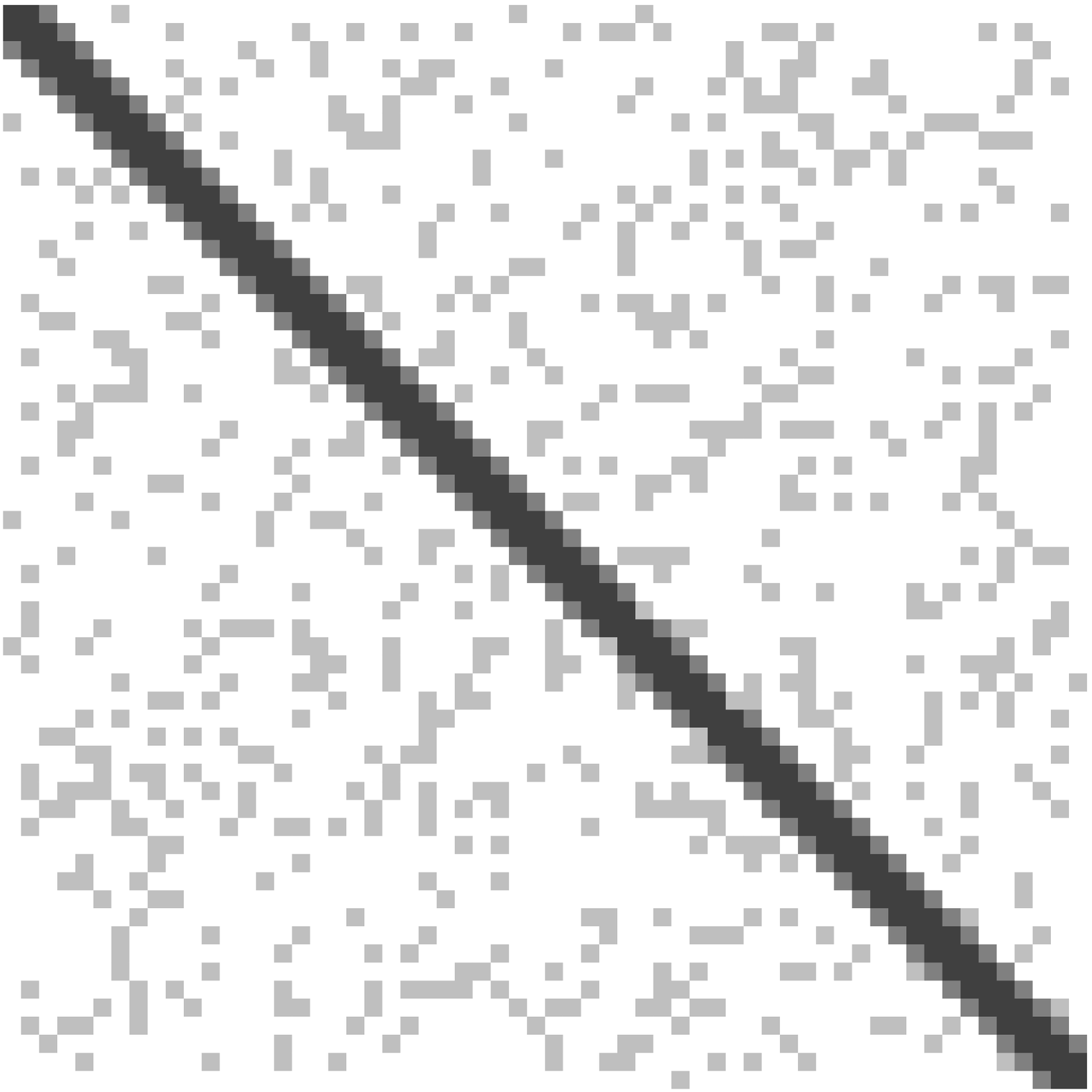} }
  \hspace*{0.06\textwidth} \subfloat[][SCAD
  ]{\includegraphics[width=0.19\textwidth]{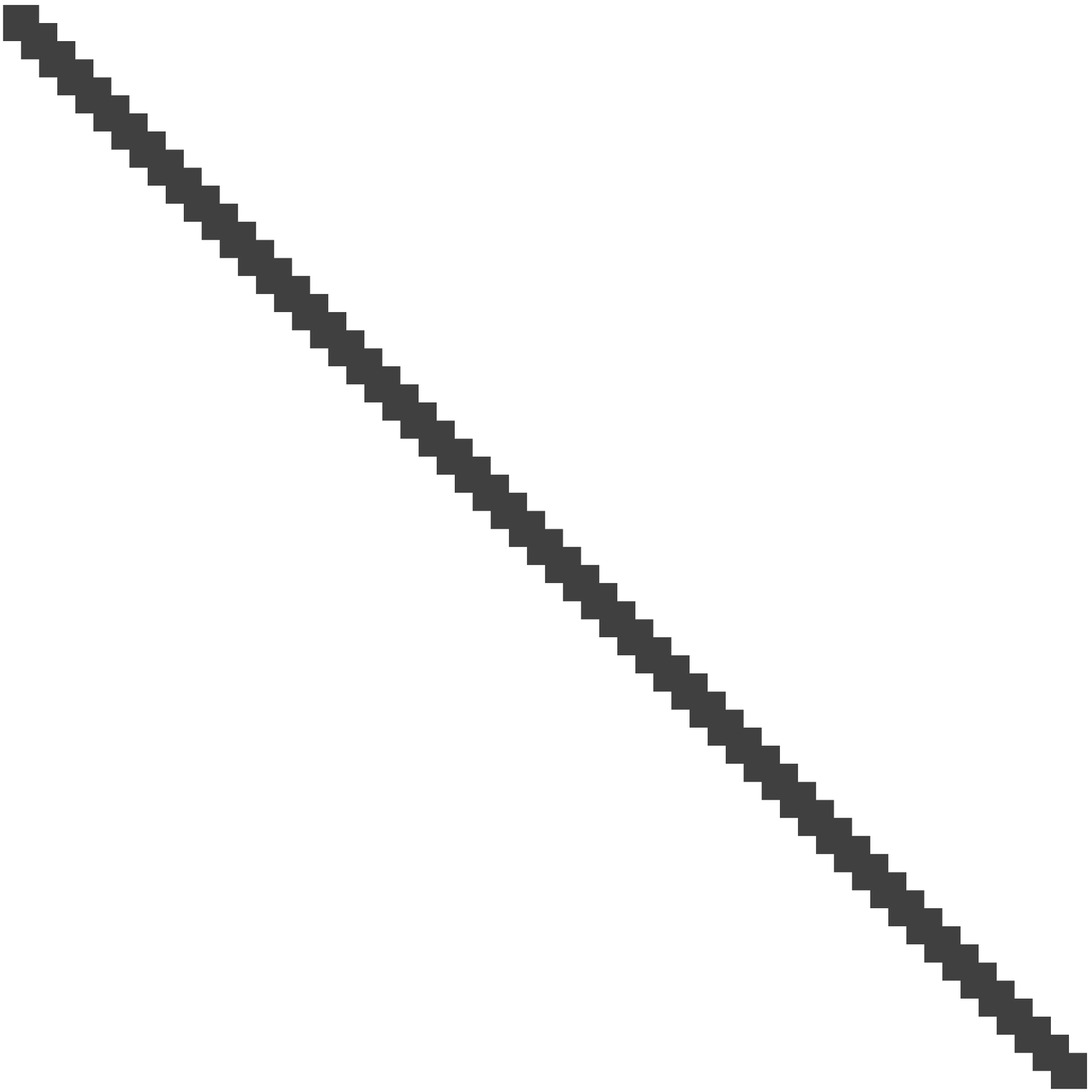} }
  \\
  \vspace*{2em}
  Model 2\\
  \subfloat[][Truth]{\includegraphics[width=0.19\textwidth]{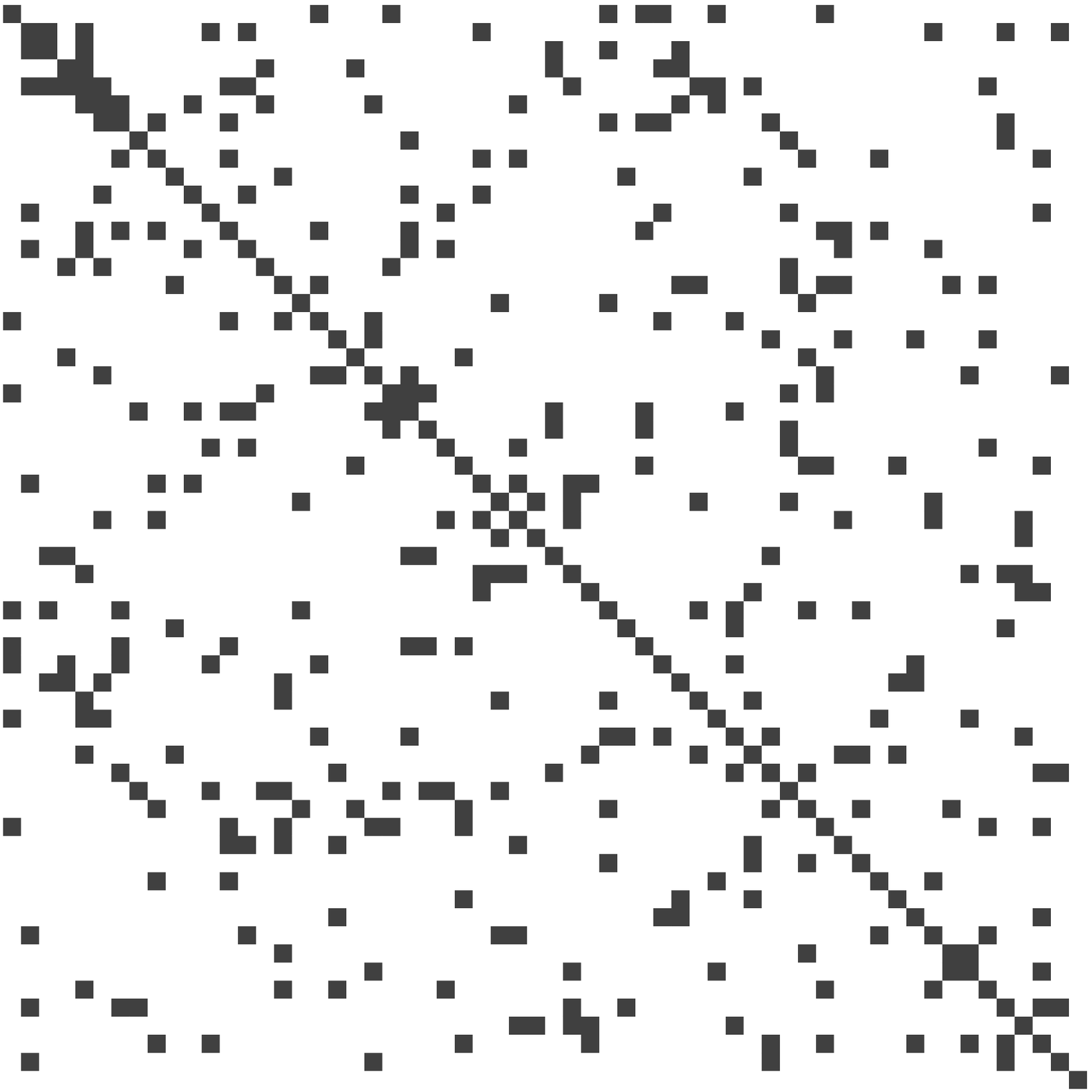}
  } \hspace*{0.06\textwidth} \subfloat[][CLIME
  ]{\includegraphics[width=0.19\textwidth]{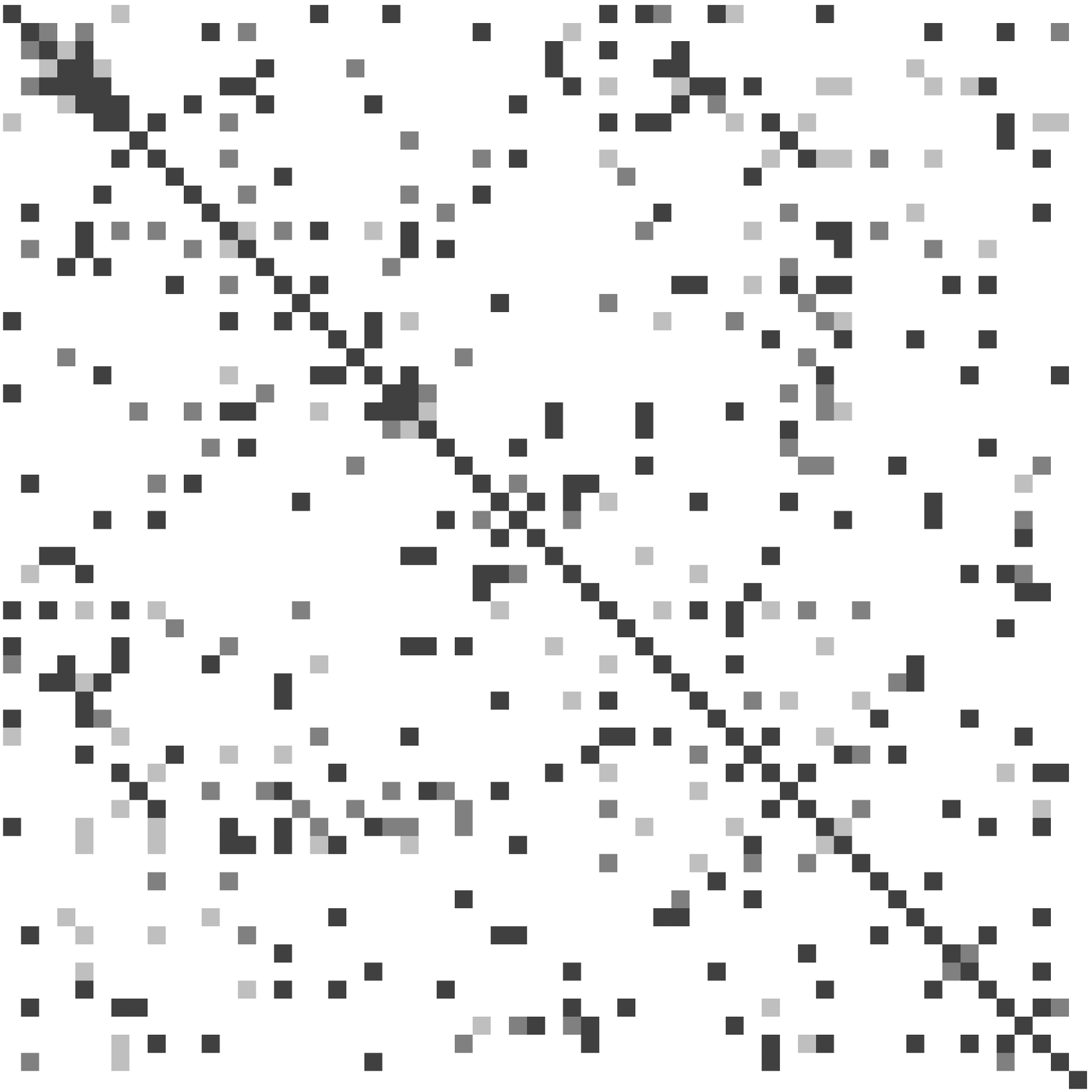} }
  \hspace*{0.06\textwidth} \subfloat[][Glasso
  ]{\includegraphics[width=0.19\textwidth]{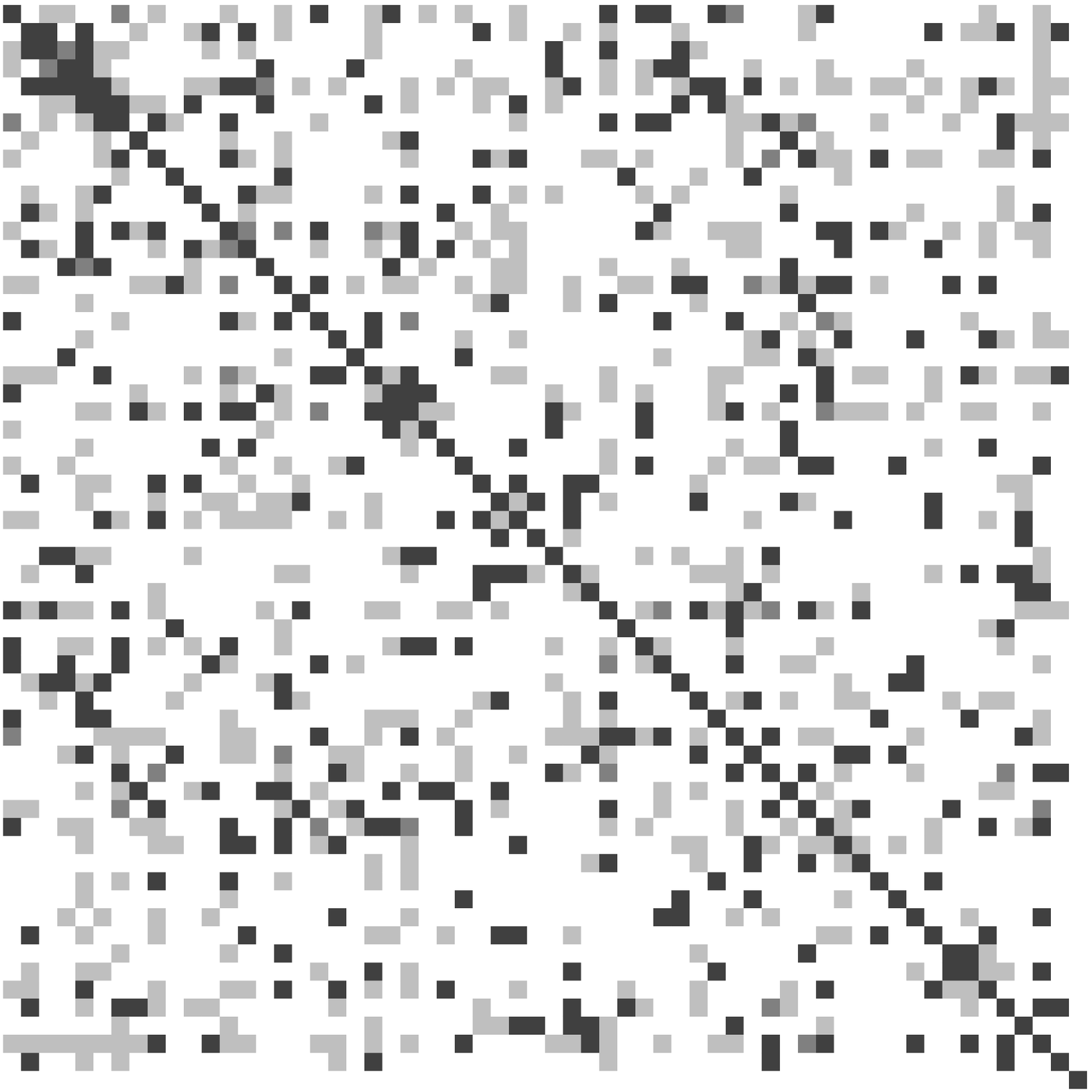} }
  \hspace*{0.06\textwidth} \subfloat[][SCAD
  ]{\includegraphics[width=0.19\textwidth]{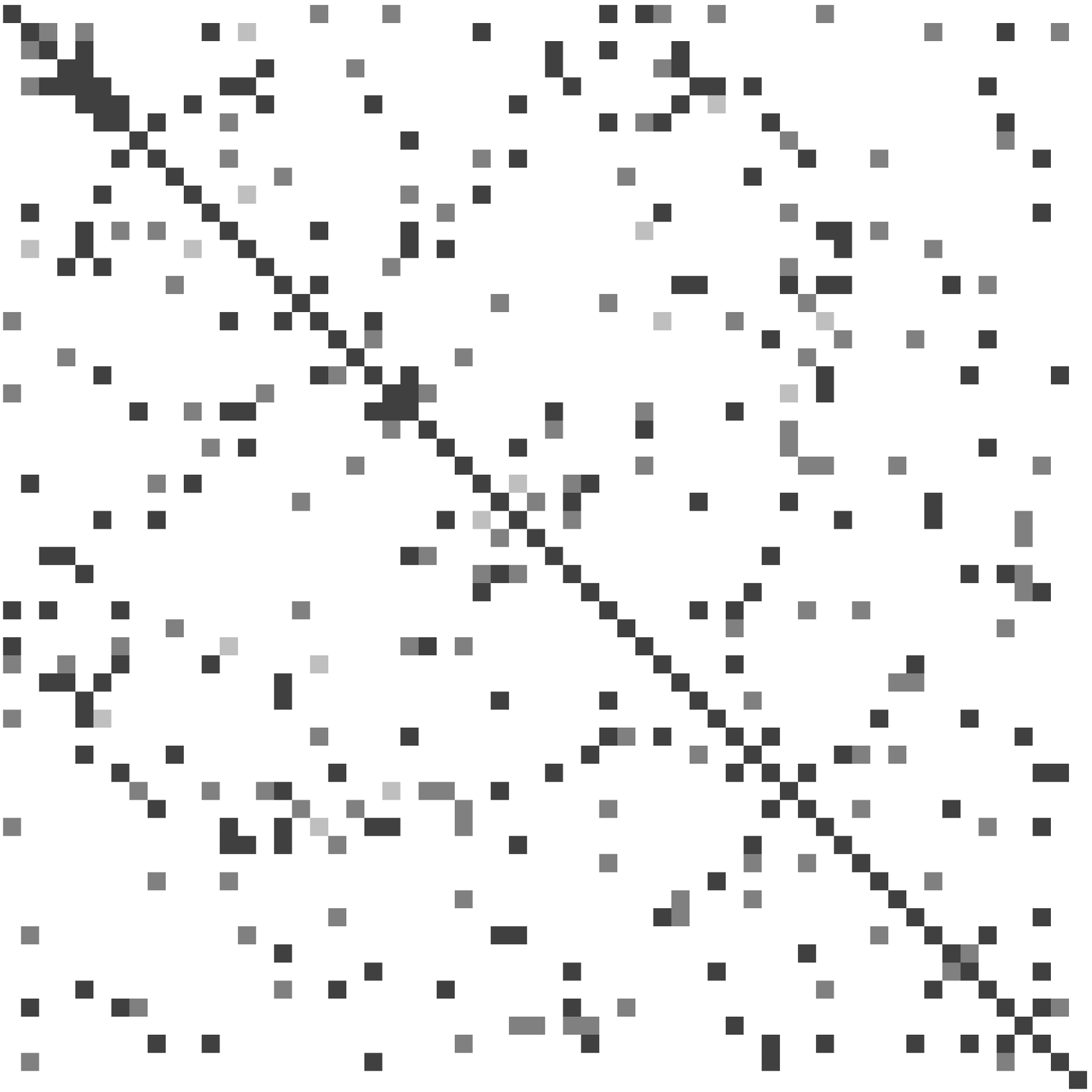} }
  \caption{Heatmaps of the frequency of the zeros identified for each
    entry of the precision matrix (when $p=60$) out of $100$
    replications.  White color is $100$ zeros identified out of $100$
    runs, and black is $0/100$.}
  \label{fig:heatmap}
\end{figure}

To better illustrate the recovery performance elementwise, the
heatmaps of the nonzeros identified out of $100$ replications are
pictured in Figure \ref{fig:heatmap}.  All the heatmaps suggest that
CLIME is more sparse than Glasso, and by visual inspection the
sparsity pattern recovered by CLIME has significantly better
resemblance to the true model than Glasso.  When the true model has
significant nonzero elements scattered on the off diagonals, Glasso
tends to include more nonzero elements than needed.  SCAD produces the
most sparse among the three but could again zero out more true nonzero
entries as shown in Model~1.  Similar patterns are observed in our
experiments for other values of $p$.

\subsection{Analysis of a breast cancer dataset}

We now apply our method CLIME on a real data example.  The breast
cancer data were analyzed by Hess et al.\ (2006) and are available at\\
\verb+http://bioinformatics.mdanderson.org/+.  The data set consists
of $22283$ gene expression levels of $133$ subjects, $34$ of which have
achieved pathological complete response (pCR) and the rest with
residual disease (RD).  The pCR subjects are considered to have high
chance of cancer-free survival in the long term, and thus it is of
great interest to study the response states of the patients (pCR or
RD) to neoadjuvant (preoperative) chemotherapy.  Based on the
estimated inverse covariance matrix of the gene expression levels, we
apply the linear discriminant analysis (LDA) to predict whether a
subject can achieve the pCR state or not.

For a fair comparison with other methods on estimating the inverse
covariance matrix, we follow the same analysis scheme discussed in Fan
et al.\ (2009) and the references therein.  For completeness, we here
give a brief description of these steps.  The data are randomly
divided into the training and the testing data sets. A stratified
sampling approach is applied to divide the data, where $5$ pCR
subjects and $16$ RD subjects are randomly selected to constitute the
testing data (roughly $1/6$ of the subjects in each group).  The
remaining subjects form the training set.  On the training set, a two
sample $t$ test is performed between the two groups for each gene, and
the $113$ most significant genes (smallest $p$-values) are retained as
the covariates for prediction.  Note that the size of the training
sample is $112$, one less than the variable size, hence it allows us
to examine the performance when $p > n$. The gene data are then
standardized by the estimated standard deviation, estimated from the
training data.  Finally, following the LDA framework, the normalized
gene expression data are assumed to be normally distributed as
$N(\boldsymbol{\mu}_k, \boldsymbol{\Sigma})$, where the two groups are assumed to have the same
covariance matrix $\boldsymbol{\Sigma}$ but different means $\boldsymbol{\mu}_k$, $k=1$ for pCR
and $k=2$ for RD.  The estimated inverse covariance $\hat{\boldsymbol{\Omega}}$
produced by different methods is used in the linear discriminant
scores
\begin{equation}
  \delta_k (\boldsymbol{x}) = \boldsymbol{x}^T \hat{\boldsymbol{\Omega}} \hat{\boldsymbol{\mu}}_k  - \frac{1}{2}
  \hat{\boldsymbol{\mu}}_k^T \hat{\boldsymbol{\Omega}} \hat{\boldsymbol{\mu}}_k  + \log \hat{\pi}_k,
\end{equation}
where $\hat{\pi}_k = n_k/n$ is the proportion of group $k$ subjects in
the training set and $\hat{\boldsymbol{\mu}}_k = (1/n_k) \sum_{i\in \text{group }k}
\boldsymbol{x}_i$ is the within-group average vector in the training set.  The
classification rule is taken to be $\hat{k}(\boldsymbol{x}) = \argmax
\delta_{k}(\boldsymbol{x})$ for $k=1,2$. 

The classification performance is clearly associated with the
estimation accuracy of $\hat{\boldsymbol{\Omega}}$.  We use the testing data set to
assess the estimation performance and compare with the existing
results in Fan et al.\ (2009) using the same criterion.  For the
tuning parameters, we use a $6$ fold cross validation on the training
data for picking $\lambda$.  The above estimation scheme is repeated
$100$ times.

To compare the classification performance, specificity, sensitivity
and Mathews Correlation Coefficient (MCC) criteria are used, which are
defined as follows:
\begin{align*}
  \text{Specificity} = \frac{\text{TN}}{\text{TN} + \text{FP}}, \quad
  \text{Sensitivity} = \frac{\text{TP}}{\text{TP} + \text{FN}},\\
  \text{MCC} = \frac{\text{TP} \times \text{TN} -\text{FP} \times
    \text{FN} }{\sqrt{(\text{TP} + \text{FP} )(\text{TP} + \text{FN}
      )(\text{TN} + \text{FP} )(\text{TN} + \text{FN} ) } },
\end{align*}
where $\text{TP}$ and $\text{TN}$ stand for true positives (pCR) and
true negatives (RD) respectively, and $\text{FP}$ and $\text{FN}$ for
false positives/negatives.  The larger the criterion value, the better
the classification performance.
The averages and standard errors of the above criteria along with
the number of nonzero entries in $\hat{\boldsymbol{\Omega}}$ over $100$
replications are reported in Table \ref{tb:pcr}.  The Glasso, Adaptive
lasso and SCAD results are taken from Fan et al.\ (2009), which uses
the same procedure on the same data set, except that we here use
$\hat{\boldsymbol{\Omega}}_{\rm CLIME}$ in place of
$\hat{\boldsymbol{\Omega}}$.  
\begin{table}[h!]
  \centering
  \caption{Comparison of average(SE) pCR
    classification errors over $100$ replications.  Glasso, Adaptive
    lasso and SCAD
    results are taken from Fan et al.\ (2009),  Table 2.}
  \begin{tabular}[htb!]{|c c c c c|}
    \hline
    Method & Specificity & Sensitivity & MCC & Nonzero entries in
    $\hat{\boldsymbol{\Omega}}$\\
    \hline
    Glasso & $0.768(0.009)$ & $0.630(0.021)$ & $0.366(0.018)$ &
    $3923(2)$ \\
    Adaptive lasso & $0.787(0.009)$ & $0.622(0.022)$ & $0.381(0.018)$ &
    $1233(1)$ \\
    SCAD & $0.794(0.009)$ & $0.634(0.022)$ & $0.402(0.020)$ &
    $674(1)$\\
    CLIME &  $0.749(0.005)$ & $0.806(0.017)$ & $0.506(0.020)$ &
    $492(7)$\\
    \hline
  \end{tabular}
  \label{tb:pcr}
\end{table}

It is clear that CLIME significantly outperforms on the sensitivity
and is comparable with other two methods on the specificity.  The
overall classification performance measured by MCC overwhelmingly
favors our method CLIME, which shows an $25\%$ improvement over the
best alternative methods.  CLIME also produced the most sparse matrix
than all other alternatives, which is usually favorable for
interpretation purposes on real data sets.

\section{Discussion}
\label{sec:conclusion}
This paper develops a new constrained $\ell_1$ minimization method for
estimating high dimensional precision matrices. Both the method and
the analysis are relatively simple and straightforward, and may be
extended to other related problems.   Moreover,
the method and the results are not restricted to a specific sparsity pattern.
Thus the estimator can be used to recover a wide class of matrices in theory as
well as in applications. In particular, when applying our method to
covariance selection in Gaussian graphical models, the theoretical
results can be established without assuming the irrepresentable condition
in Ravikumar et al.\ (2008), which is very stringent and hard to check
in practice.

Several papers, such as Yuan and Lin (2007), Rothman et al.\  (2008)
and Ravikumar et al.\ (2008), estimate the precision matrix by solving
the optimization problem (\ref{lass}) with $\ell_1$ penalty only on
the off diagonal entries, which is slightly different from our starting
point (\ref{glass}) presented here.  One can also similarly considered
the following optimization problem
\begin{eqnarray*}
  \min\|\boldsymbol{\Omega}\|_{1,\rm off} ~~\mbox{subj}~~
  |\boldsymbol{\Sigma}_{n}\boldsymbol{\Omega}-\boldsymbol{I}|_{\infty}\leq \lambda_{n},~~\boldsymbol{\Omega}\in
  \RR^{p\times p}.
\end{eqnarray*}
Analogous results can also be established for the above estimator. We
omit them in this paper, due to high resemblance in proof techniques
and conclusions.

There are several possible extensions for our method. For example,
Zhou et al.\ (2008) considered the time varying undirected graphs and
estimated $\boldsymbol{\Sigma}(t)^{-1}$ by Glasso. It would be very interesting to
study the estimation of $\boldsymbol{\Sigma}(t)^{-1}$ by our method.  Ravikumar and
Wainwright (2009) considered high-dimensional Ising model selection
using $\ell_{1}$-regularized logistic regression. It would be interesting
to apply our method to their setting as well.

Another important subject is to investigate the theoretical property
of the tuning parameter selected by cross-validation method, though
from our experiments CLIME is not very sensitive to the
choice of the tuning parameter.  An example of such results on cross
validation can be found in Bickel and Levina (2008b) on thresholding.

After this paper was submitted, it came to our attention that Zhang
(2010) proposed a precision matrix estimator, called GMACS,
which is the solution of the following optimization problem:
\begin{equation*}
  \min\|\boldsymbol{\Omega}\|_{L_1} ~~\mbox{subject to:}~~ |\boldsymbol{\Sigma}_{n}\boldsymbol{\Omega}-\boldsymbol{I}|_{\infty}\leq
  \lambda_{n},~~\boldsymbol{\Omega}\in \RR^{p\times p}. 
\end{equation*}
The objective function here is different from that of CLIME, and this
basic version cannot be solved column by column and is not as easy to implement.  Zhang (2010)
considers only the Gaussian case and $\ell_0$ balls, whereas we consider
subgaussian and polynomial-tail distributions and more general $\ell_q$ balls.  
Also, the GMACS estimator requires an additional thresholding step in order for
the rates to hold over $\ell_0$ balls.  In contrast, CLIME
does not need an additional thresholding step and the rates hold over general $\ell_q$ balls.

\section{Proof of Main Results}
\label{sec:proof}

{\bf Proof of Lemma \ref{le1}.}  Write
$\boldsymbol{\Omega}=(\boldsymbol{\omega}_{1},\dotsc,\boldsymbol{\omega}_{p})$, where $\boldsymbol{\omega}_{i}\in
\RR^{p}$. The constraint $|\boldsymbol{\Sigma}_{n}\boldsymbol{\Omega}-\boldsymbol{I}|_{\infty}\leq
\lambda_{n}$ is equivalent to
\begin{eqnarray*}
  |\boldsymbol{\Sigma}_{n}\boldsymbol{\omega}_{i}-\boldsymbol{e}_{i}|_{\infty}\leq
  \lambda_{n}~~\mbox{for~~}1\leq i\leq p.
\end{eqnarray*}
Thus we have
\begin{eqnarray}\label{c2}
  |\hat{\boldsymbol{\omega}}^{1}_{i}|_{1}\geq |\hat{\boldsymbol{\beta}}_{i}|_{1}~~\mbox{for
    $1\leq i\leq p$.}
\end{eqnarray}
Since $|\boldsymbol{\Sigma}_{n}\hat{\textbf{B}}-\boldsymbol{I}|_{\infty}\leq \lambda_{n}$, by
the definitions of $\{\hat{\boldsymbol{\Omega}}_{1}\}$, we have
\begin{eqnarray}\label{c3}
  \|\hat{\boldsymbol{\Omega}}_{1}\|_{1}\leq \|\hat{\textbf{B}}\|_{1}.
\end{eqnarray}
By(\ref{c2}) and (\ref{c3}), we have $\hat{\textbf{B}}\in
\{\hat{\boldsymbol{\Omega}}_{1}\}$. On the other hand, if $\hat{\boldsymbol{\Omega}}_{1}\notin
\{\hat{\textbf{B}}\}$, then there exists an $i$ such that
$|\hat{\boldsymbol{\omega}}_{i}|_{1}>|\hat{\boldsymbol{\beta}}_{i}|_{1}$. Hence by (\ref{c2}) we
have $\|\hat{\boldsymbol{\Omega}}_{1}\|_{1}>\|\hat{\textbf{B}}\|_{1}$. This is in
conflict with (\ref{c3}).\qed
\\

The main results all rely on Theorem \ref{th1}, which upper bounds the
elementwise $\ell_\infty$ norm.  We will prove it first.

\noindent{\bf Proof of Theorem \ref{th1}.} Let $\hat{\boldsymbol{\beta}}_{i,\rho}$
be a solution of (\ref{o1}) by replacing $\boldsymbol{\Sigma}_{n}$ with
$\boldsymbol{\Sigma}_{n,\rho}$. Note that Lemma \ref{le1} still holds for
$\hat{\boldsymbol{\Omega}}_{n,\rho}$ and $\{\hat{\boldsymbol{\beta}}_{i,\rho}\}$ with $\rho\geq
0$. For notation briefness, we only prove the theorem for
$\rho=0$. The proof is exactly the same for general $\rho>0$. By the
condition in Theorem \ref{th1},
\begin{eqnarray}\label{c4}
  |\boldsymbol{\Sigma}_{0}-\boldsymbol{\Sigma}_{n}|_{\infty}\leq
  \lambda_{n}/\|\boldsymbol{\Omega}_{0}\|_{L_{1}}.
\end{eqnarray}
Then we have
\begin{eqnarray}\label{p1}
  |\boldsymbol{I}-\boldsymbol{\Sigma}_{n}\boldsymbol{\Omega}_{0}|_{\infty}=
  |(\boldsymbol{\Sigma}_{0}-\boldsymbol{\Sigma}_{n})\boldsymbol{\Omega}_{0}|_{\infty}\leq
  \|\boldsymbol{\Omega}_{0}\|_{L_{1}}|\boldsymbol{\Sigma}_{0}-\boldsymbol{\Sigma}_{n}|_{\infty}\leq
  \lambda_{n},
\end{eqnarray}
where we used the inequality $|\boldsymbol{A} \boldsymbol{B}|_{\infty}\leq |\boldsymbol{A}|_{\infty}\|\boldsymbol{B}\|_{L_{1}}$
for matrices $\boldsymbol{A}, \boldsymbol{B}$ of appropriate sizes.  By the definition of
$\hat{\boldsymbol{\beta}}_{i}$, we can see that $|\hat{\boldsymbol{\beta}}_{i}|_{1}\leq
\|\boldsymbol{\Omega}_{0}\|_{L_{1}}$ for $1\leq i\leq p$. By Lemma \ref{le1},
\begin{eqnarray}\label{a8}
  \|\hat{\boldsymbol{\Omega}}_{1}\|_{L_{1}}\leq \|\boldsymbol{\Omega}_{0}\|_{L_{1}}.
\end{eqnarray}
We have
\begin{eqnarray}\label{a9}
  |\boldsymbol{\Sigma}_{n}(\hat{\boldsymbol{\Omega}}_{1}-\boldsymbol{\Omega}_{0})|_{\infty}\leq
  |\boldsymbol{\Sigma}_{n}\hat{\boldsymbol{\Omega}}_{1}-\boldsymbol{I}|_{\infty}+|\boldsymbol{I}-\boldsymbol{\Sigma}_{n}\boldsymbol{\Omega}_{0}|_{\infty}\leq
  2\lambda_{n}.
\end{eqnarray}
Therefore by (\ref{c4})-(\ref{a9}),
\begin{eqnarray*}
  |\boldsymbol{\Sigma}_{0}(\hat{\boldsymbol{\Omega}}_{1}-\boldsymbol{\Omega}_{0})|_{\infty}&\leq&|\boldsymbol{\Sigma}_{n}(\hat{\boldsymbol{\Omega}}_{1}-\boldsymbol{\Omega}_{0})|_{\infty}+
  |(\boldsymbol{\Sigma}_{n}-\boldsymbol{\Sigma}_{0})(\hat{\boldsymbol{\Omega}}_{1}-\boldsymbol{\Omega}_{0})|_{\infty}\cr
  &\leq&
  2\lambda_{n}+\|\hat{\boldsymbol{\Omega}}_{1}-\boldsymbol{\Omega}_{0}\|_{L_{1}}|\boldsymbol{\Sigma}_{n}-\boldsymbol{\Sigma}_{0}|_{\infty}\leq
  4\lambda_{n}.
\end{eqnarray*}
It follows that
\begin{eqnarray*}
  |\hat{\boldsymbol{\Omega}}_{1}-\boldsymbol{\Omega}_{0}|_{\infty}\leq
  \|\boldsymbol{\Omega}_{0}\|_{L_{1}}|\boldsymbol{\Sigma}_{0}(\hat{\boldsymbol{\Omega}}_{1}-\boldsymbol{\Omega}_{0})|_{\infty}\leq
  4\|\boldsymbol{\Omega}_{0}\|_{L_{1}}\lambda_{n}.
\end{eqnarray*}
This establishes (\ref{t2}) by the definition in \eqref{me1}.

We next prove (\ref{t1}). Let
$t_{n}=|\hat{\boldsymbol{\Omega}}-\boldsymbol{\Omega}_{0}|_{\infty}$ and define
\begin{eqnarray*}
  &&\boldsymbol{h}_{j}=\hat{\boldsymbol{\omega}}_{j}-\boldsymbol{\omega}^{0}_{j},\cr
  && \boldsymbol{h}^{1}_{j}=(\hat{\boldsymbol{\omega}}_{ij}I\{|\hat{{{\omega}}}_{ij}|\geq
  2t_{n}\};1\leq i\leq
  p)^{T}-\boldsymbol{\omega}^{0}_{j},~~\boldsymbol{h}^{2}_{j}=\boldsymbol{h}_{j}-\boldsymbol{h}^{1}_{j}.
\end{eqnarray*}
By the definition (\ref{me1}) of $\hat{\boldsymbol{\Omega}}$, we have
$|\hat{\boldsymbol{\omega}}_{j}|_{1}\leq |\hat{\boldsymbol{\omega}}^{1}_{j}|_{1}\leq
|\boldsymbol{\omega}^{0}_{j}|_{1}$. Then
\begin{eqnarray*}
  |\boldsymbol{\omega}^{0}_{j}|_{1}-|\boldsymbol{h}^{1}_{j}|_{1}+|\boldsymbol{h}^{2}_{j}|_{1} \leq
  |\boldsymbol{\omega}^{0}_{j}+\boldsymbol{h}^{1}_{j}|_{1}+|\boldsymbol{h}^{2}_{j}|_{1}=|\hat{\boldsymbol{\omega}}_{j}|_{1}\leq
  |\boldsymbol{\omega}^{0}_{j}|_{1},
\end{eqnarray*}
which implies that $|\boldsymbol{h}^{2}_{j}|_{1}\leq |\boldsymbol{h}^{1}_{j}|_{1}$. This follows
that $|\boldsymbol{h}_{j}|_{1}\leq 2|\boldsymbol{h}^{1}_{j}|_{1}$. So we only need to upper
bound $|\boldsymbol{h}^{1}_{j}|_{1}$. We have
\begin{eqnarray}\label{t4}
  |\boldsymbol{h}^{1}_{j}|_{1}&=&\sum_{i=1}^{p}|\hat{\omega}_{ij}I\{|\hat{\omega}_{ij}|\geq
  2t_{n}\}-\omega^{0}_{ij}|\cr &\leq
  &\sum_{i=1}^{p}|\omega^{0}_{ij}I\{|\omega^{0}_{ij}|\leq
  2t_{n}\}|+\sum_{i=1}^{p}|\hat{\omega}_{ij}I\{|\hat{\omega}_{ij}|\geq
  2t_{n}\}-\omega^{0}_{ij}I\{|\omega^{0}_{ij}|\geq 2t_{n}\}|\cr &\leq
  &(2t_{n})^{1-q}s_{0}(p)+t_{n}\sum_{i=1}^{p}I\{|\hat{\omega}_{ij}|\geq
  2t_{n}\}+\sum_{i=1}^{p}|\omega^{0}_{ij}||I\{|\hat{\omega}_{ij}|\geq
  2t_{n}\}-I\{|\omega^{0}_{ij}|\geq 2t_{n}\}|\cr &\leq &
  (2t_{n})^{1-q}s_{0}(p)+t_{n}\sum_{i=1}^{p}I\{|\omega^{0}_{ij}|\geq
  t_{n}\}+\sum_{i=1}^{p}|\omega^{0}_{ij}|I\{||\omega^{0}_{ij}|-2t_{n}|\leq
  |\hat{\omega}_{ij}-\omega^{0}_{ij}|\}\cr &\leq
  &(2t_{n})^{1-q}s_{0}(p)+(t_{n})^{1-q}s_{0}(p)+(3t_{n})^{1-q}s_{0}(p)\cr
  &\leq &(1+2^{1-q}+3^{1-q})t_{n}^{1-q}s_{0}(p),
\end{eqnarray}
where we used the following inequality: for any $a,b,c\in \RR$, we
have
\begin{eqnarray*}
  |I\{a<c\}-I\{b<c\}|\leq I\{|b-c|< |a-b|\}.
\end{eqnarray*}
This completes the proof of (\ref{t1}).

Finally, (\ref{t3}) follows from (\ref{t4}), (\ref{t2}) and the
inequality $\|\boldsymbol{A}\|^{2}_{F}\leq p\|\boldsymbol{A}\|_{L_{1}}|\boldsymbol{A}|_{\infty}$ for any
$p\times p$ matrix.\qed
\\

\noindent{\bf Proof of Theorems  \ref{thn-2} (i) and \ref{cr1} (i).}
By Theorem \ref{th1}, we only need to prove
\begin{eqnarray}\label{c9}
  \max_{ij}|\hat{\sigma}_{ij}-\sigma^{0}_{ij}|\leq C_{0}\sqrt{\log
    p/n}
\end{eqnarray}
with probability greater than $1-4p^{-\tau}$ under (C1). Without loss
of generality, we assume $\ep \textbf{X}=0$. Let
$\boldsymbol{\Sigma}^{0}_{n}:=n^{-1}\sum_{k=1}^{n}\textbf{X}_{k}\textbf{X}_{k}^{T}$
and $Y_{kij}=X_{ki}X_{kj}-\ep X_{ki}X_{kj}$.  Then we have
$\boldsymbol{\Sigma}_{n}=\boldsymbol{\Sigma}^{0}_{n}-\bar{\textbf{X}}\bar{\textbf{X}}^{T}$.  Let
$t=\eta\sqrt{\log p/n}$. Using the inequality $|e^{s}-1-s|\leq
s^{2}e^{\max(s,0)}$ for any $s\in R$ and letting
$C_{K1}=2+\tau+\eta^{-1}K^{2}$, by basic calculations, we can get
\begin{eqnarray*} \pr\Big{(} \sum_{k=1}^{n}Y_{kij}\geq
  \eta^{-1}C_{K1}\sqrt{n\log p}\Big{)} &\leq& e^{-C_{K1}\log
    p}\Big{(}\ep\exp(tY_{kij})\Big{)}^{n}\cr &\leq
  &\exp\Big{(}-C_{K1}\log p+nt^{2}\ep
  Y^{2}_{kij}e^{t|Y_{kij}|}\Big{)}\cr &\leq & \exp\Big{(}-C_{K1}\log
  p+\eta^{-1}K^{2}\log p\Big{)}\cr &\leq &\exp(-(\tau+2)\log p).
\end{eqnarray*}
Hence we have
\begin{eqnarray}\label{c5}
  \pr\Big{(}|\boldsymbol{\Sigma}^{0}_{n}-\boldsymbol{\Sigma}_{0}|_{\infty}\geq
  \eta^{-1}C_{K1}\sqrt{\log p/ n}\Big{)}\leq 2p^{-\tau}.
\end{eqnarray}
By the simple inequality $e^{s}\leq e^{s^{2}+1}$ for $s>0$, we have
$\ep e^{t|X_{j}|}\leq eK$ for all $t\leq \eta^{1/2}$. Let $C_{K2}=
2+\tau+\eta^{-1}e^{2}K^{2}$ and $a_{n}=C^{2}_{K2}(\log p/n)^{1/2}$.
As above, we can show that
\begin{eqnarray}\label{c6}
  \pr\Big{(}|\bar{\textbf{X}}\bar{\textbf{X}}^{T}|_{\infty}\geq
  \eta^{-2}a_{n}\sqrt{\log p/n}\Big{)}&\leq& p\max_{i}\pr\Big{(}
  \sum_{k=1}^{n}X_{ki}\geq \eta^{-1}C_{K2}\sqrt{n\log p}\Big{)}\cr &
  &+p\max_{i}\pr\Big{(} -\sum_{k=1}^{n}X_{ki}\geq
  \eta^{-1}C_{K2}\sqrt{n\log p}\Big{)}\cr
  &\leq&
  2p^{-\tau-1}.
\end{eqnarray}
By (\ref{c5}), (\ref{c6}) and the inequality
$C_{0}>\eta^{-1}C_{K1}+\eta^{-2}a_{n}$, we see that (\ref{c9})
holds.\qed
\\

\noindent{\bf Proof of Theorems  \ref{thn-2} (ii) and \ref{cr1}
(ii).}
Let
\begin{eqnarray*}
  &&\bar{Y}_{kij}=X_{ki}X_{kj}I\{|X_{ki}X_{kj}|\leq \sqrt{n/(\log
    p)^{3}}\}-\ep X_{ki}X_{kj}I\{|X_{ki}X_{kj}|\leq \sqrt{n/(\log
    p)^{3}}\},\cr &&\check{Y}_{kij}=Y_{kij}-\bar{Y}_{kij}.
\end{eqnarray*}
Since $b_{n}:=\max_{i,j}\ep |X_{ki}X_{kj}|I\{|X_{ki}X_{kj}|\geq
\sqrt{n/(\log p)^{3}}\}=O(1)n^{-\gamma-1/2}$, we have by (C2),
\begin{eqnarray*}
  &&\pr\Big{(}\max_{i,j}|\sum_{k=1}^{n}\check{Y}_{kij}|\geq
  2nb_{n}\Big{)}\cr &&\leq
  \pr\Big{(}\max_{i,j}|\sum_{k=1}^{n}X_{ki}X_{kj}I\{|X_{ki}X_{kj}|>
  \sqrt{n/(\log p)^{3}}\}|\geq nb_{n}\Big{)}\cr && \leq
  \pr\Big{(}\max_{i,j}\sum_{k=1}^{n}|X_{ki}X_{kj}|I\{X^{2}_{ki}+X^{2}_{kj}\geq
  2\sqrt{n/(\log p)^{3}}\}\geq nb_{n}\Big{)}\cr &&\leq
  \pr\Big{(}\max_{k,i}X_{ki}^{2}\geq \sqrt{n/(\log p)^{3}}\Big{)}\cr
  &&\leq
  pn\pr\Big{(}X^{2}_{1}\geq \sqrt{n/(\log p)^{3}}\Big{)}\cr &&=
  O(1)n^{-\delta/8}.
\end{eqnarray*}
By Bernstein's inequality (cf. Bennett (1962)) and some elementary
calculations,
\begin{eqnarray*}
  &&\pr\Big{(}\max_{i,j}|\sum_{k=1}^{n}\bar{Y}_{kij}|\geq
  \sqrt{(\theta+1)(4+\tau) n\log p}\Big{)}\cr &&\leq p^{2}\max_{i,j}
  \pr\Big{(}|\sum_{k=1}^{n}\bar{Y}_{kij}|\geq \sqrt{(\theta+1) (4+\tau)
    n\log p}\Big{)}\cr &&\leq 2p^{2} \max_{i,j} \exp\Big{(}-\frac{(\theta+1)(4+\tau)
    n\log p}{2n\ep \bar{Y}^{2}_{1ij}+\sqrt{(\theta+1) (64+16\tau)} n/(3\log
    p)}\Big{)}\cr &&=O(1) p^{-\tau/2}.
\end{eqnarray*}
So we have
\begin{eqnarray}\label{c13}
  \pr\Big{(}|\boldsymbol{\Sigma}^{0}_{n}-\boldsymbol{\Sigma}_{0}|_{\infty}\geq
  \sqrt{(\theta+1) (4+\tau) \log
    p/n}+2b_{n}\Big{)}=O\Big{(}n^{-\delta/8}+p^{-\tau/2}\Big{)}.
\end{eqnarray}
Using the same truncation argument and Bernstein's inequality, we can
show that
\begin{eqnarray*}
  \pr\Big{(} \max_{i}\Big{|}\sum_{k=1}^{n}X_{ki}\Big{|}\geq
  \sqrt{\max_{i}\sigma^{0}_{ii}(4+\tau)n\log
    p}\Big{)}=O\Big{(}n^{-\delta/8}+p^{-\tau/2}\Big{)}.
\end{eqnarray*}
Hence
\begin{eqnarray}\label{c12}
  \pr\Big{(}|\bar{\textbf{X}}\bar{\textbf{X}}^{T}|_{\infty}\geq
  \max_{i}\sigma^{0}_{ii}(4+\tau)\log
  p/n\Big{)}=O\Big{(}n^{-\delta/8}+p^{-\tau/2}\Big{)}.
\end{eqnarray}
Combining (\ref{c13}) and (\ref{c12}), we have
\begin{eqnarray}\label{c11}
  \max_{ij}|\hat{\sigma}_{ij}-\sigma^{0}_{ij}|\leq
  \sqrt{(\theta+1) (5+\tau) \log p/n}
\end{eqnarray}
with probability greater than
$1-O\Big{(}n^{-\delta/8}+p^{-\tau/2}\Big{)}$.  The proof is completed
by \eqref{c11} and Theorem \ref{th1}. \qed
\\

\noindent{\bf Proof of Theorems \ref{thn-3} and \ref{n-cr1}.} Since
$\boldsymbol{\Sigma}^{-1}_{n,\rho}$ is a feasible point, we have by (\ref{a10}),
\begin{eqnarray*}
  \|\hat{\boldsymbol{\Omega}}_{\rho}\|_{1}\leq\|\hat{\boldsymbol{\Omega}}_{1\rho}\|_{1}\leq
  \|\boldsymbol{\Sigma}^{-1}_{n,\rho}\|_{1}\leq p^{2}\max(\sqrt{\frac{n}{\log
      p}},p^{\alpha}).
\end{eqnarray*}
By (\ref{c9}), Theorem \ref{th1}, the fact $p\geq n^{\xi}$ and since $\tau$ is large enough,
 we have
\begin{eqnarray*}
  \sup_{\boldsymbol{\Omega}_{0}\in\mathcal{U}}\ep\|\hat{\boldsymbol{\Omega}}_{\rho}-\boldsymbol{\Omega}_{0}\|^{2}_{2}
  &=&\sup_{\boldsymbol{\Omega}_{0}\in\mathcal{U}}\ep\|\hat{\boldsymbol{\Omega}}_{\rho}-\boldsymbol{\Omega}_{0}\|^{2}_{2}I\{\max_{ij}|\hat{\sigma}_{ij}-\sigma^{0}_{ij}|+\rho\leq
  C_{0}\sqrt{\log p/n}\}\cr &
  &+\sup_{\boldsymbol{\Omega}_{0}\in\mathcal{U}}\ep\|\hat{\boldsymbol{\Omega}}_{\rho}-\boldsymbol{\Omega}_{0}\|^{2}_{2}I\{\max_{ij}|\hat{\sigma}_{ij}-\sigma^{0}_{ij}|+\rho>
  C_{0}\sqrt{\log p/n}\}\cr &=
  &O\Big{(}M^{4-4q}s^{2}_{0}(p)\Big{(}\frac{\log
    p}{n}\Big{)}^{1-q}\Big{)}+O\Big{(}p^{4}\max(\frac{n}{\log
    p},p^{2\alpha})p^{-\tau/2}\Big{)}\cr
  &=&O\Big{(}M^{4-4q}s^{2}_{0}(p)\Big{(}\frac{\log
    p}{n}\Big{)}^{1-q}\Big{)}.
\end{eqnarray*}
This proves Theorem \ref{thn-3}. The proof of Theorem \ref{n-cr1} is
similar.\qed
\\

\noindent{\bf Proof of Theorem \ref{th2}.} Let $k_{n}$ be an integer
satisfying $1\leq k_{n}\leq n$. Define
\begin{eqnarray*}
  &&\boldsymbol{h}_{j}=\hat{\boldsymbol{\omega}}_{j}-\boldsymbol{\omega}^{0}_{j},\cr
  && \boldsymbol{h}^{1}_{j}=(\hat{\omega}_{ij}I\{1\leq i\leq k_{n}\};1\leq i\leq
  p)^{T}-\boldsymbol{\omega}^{0}_{j},~~\boldsymbol{h}^{2}_{j}=\boldsymbol{h}_{j}-\boldsymbol{h}^{1}_{j}.
\end{eqnarray*}
By the proof of Theorem \ref{th1}, we can show that $|\boldsymbol{h}_{j}|_{1}\leq
2|\boldsymbol{h}^{1}_{j}|_{1}$. Since $\boldsymbol{\Omega}_{0}\in\mathcal{U}_{o}(\alpha,M)$, we
have $\sum_{j\geq k_{n}}|\omega^{0}_{ij}|\leq Mk^{-\alpha}_{n}$.  By
Theorem \ref{cr1},
$\sum_{j=1}^{k_{n}}|\hat{\omega}_{ij}-\omega^{0}_{ij}|=O\Big{(}k_{n}\sqrt{\log
  p/n}\Big{)}$ with probability greater than
$1-O\Big{(}n^{-\delta/8}+p^{-\tau/2}\Big{)}$. Theorem \ref{th2} (i) is
proved by taking $k_{n}=[(n/\log p)^{1/(2\alpha+2)}]$. The proof of
Theorem \ref{th2} (ii) is similar as that of Theorem \ref{thn-3}.\qed

\section*{Acknowledgment}

We would like to thank the Associate Editor and two referees for their
very helpful comments which have led to a better presentation of the paper.

\end{document}